\documentclass{article}
\pdfoutput=1
\usepackage{amsfonts}
\usepackage{amsmath}
\usepackage{amssymb}
\usepackage{color}
\usepackage{natbib}
\usepackage{booktabs}
\usepackage{multirow}
\usepackage{float}
\usepackage{natbib}
\usepackage{bm}
\usepackage{siunitx}
\usepackage{subfigure}

\usepackage{graphicx}
\usepackage{pdflscape}

\bibpunct{(}{)}{;}{a}{,}{,}

\newtheorem{theorem}{Theorem}

\newtheorem{condition}[theorem]{Condition}

\newtheorem{lemma}[theorem]{Lemma}

\newtheorem{proposition}[theorem]{Proposition}

\begin{document}

\title{Inference of Sample Complier Average Causal Effects under Experiments with Completely Randomized Design and Computer Assisted
Balance-Improving Designs}
\author{Zhen Zhong\\
		Yau Mathematical Science Center, Tsinghua University\\
		\\
		Per Johansson\\
		Department of Statistics, Uppsala
		University\\
		and Yau Mathematical Science Center,
		Tsinghua University\\
		\\
		Junni L. Zhang*\\
		National School of Development and Center for Statistical Science, 
		\\Peking University}
\date{September 30, 2023}
\maketitle

\begin{abstract}
Non-compliance is common in real world experiments.  We focus on inference about the sample complier average causal effect, that is, the average treatment effect for experimental units who are compliers.
We present three types of inference strategies for the sample complier average causal effect: the Wald estimator, regression adjustment estimators and model-based Bayesian inference.  Because modern computer assisted experimental designs have been used to improve covariate balance over complete randomization, we discuss inference under both complete randomization and a specific computer assisted experimental design -- Mahalanobis distance based rerandomization, under which asymptotic properties of the Wald estimator and regression adjustment estimators can be derived.  We use Monte Carlo simulation to compare the finite sample performance of the methods under both experimental designs. We find that under either design, the Bayesian method performs the best because it is stable, it yields smallest median absolute error and smallest median interval length.  The improvement by the Bayesian method is especially large when the fraction of compliers is small.  We present an application to a job training experiment with
non-compliance.
\end{abstract}

\section{Introduction}
One (out of many) requirements for the inference to an average treatment effect from a randomized experiment to be valid is that all experimental units comply to their treatment assignments. In practice, incomplete compliance to the assigned treatment is common. 

One standard approach is to ignore the information on compliance behavior and to focus on the intention-to-treat (ITT) analysis.  It is argued that the ITT-effect may be most policy-relevant, since one cannot in general force people to take a treatment.  However, the ITT-effect may be misleading, for example, when a treatment may appear more effective simply because subjects adhere to it to a greater extent.  The international guidelines for good clinical practice suggest that the analysis of negative side effects should be according to treatment received.\footnote{``ICH GCP guidelines (1999). ICH GCP is short for International Conference on Harmonization of Good Clinical Practice} Yet the ``as-treated'' analysis that compares those who receive treatment with those who receive control generally yields a biased estimate of the treatment effect, because the treatment and control groups are no longer similar.

When information on treatment received is available, which often is the case, the effect for those who comply to the assigned treatment can be identified under certain conditions, and scholars have argued that the complier average causal effect (CACE) ought to be the main causal estimand (see e.g. \cite{McNamee_2009,Shrier_etal_2014,Shrier_etal_2017,Steele_etal_2015}).  In the literature, often the experimental units are regarded as being randomly sampled from a superpopulation, and inference is made about the population CACE.  We instead focus only on the finite population consisting of the experimental units, and are interested in inference about the sample CACE.  This is of interest, for instance, when the units are volunteers who are different from the general population of interest.

The only previous reference that discusses inference about sample CACE and that we know of is \cite{li2017general}. They consider a linear instrumental variable model, and gives four possible forms of the confidence sets for sample CACE, including empty set, a finite closed interval, the whole real line, and a union of two infinite half-open intervals.  However, the linear instrumental variable model is overly restrictive, and the four possible forms of confidence sets are not handy for practical use.  We will consider more general settings, and construct interval estimates in the form of finite closed intervals that are familiar to practitioners.

To improve covariate balance over a complete randomization, a number of methods
have been proposed recently to utilize modern computational capabilities to find
allocations with balance in observed covariates (e.g. \cite{morgan2012rerandomization, Bertsimas_etal_2015, Kallus_2018, Lauretto_2017, Kriegeretal_2019, Kapelneretal_2021, Johansson_Schultzberg_2020,Johansson_Schultzberg_2022}).  We consider a specific computer assisted experimental design -- Mahalanobis distance based rerandomization, which is amenable to asymptotic inferences for commonly used estimators (e.g., see \citep{li2018asymptotic} and \citep{li2020rerandomization}).  When the sample average treatment effect is of interest, \cite{Zhang_Johansson_2022} proposed model-based Bayesian inference as a general strategy for inference in the computer assisted designs. 

In this paper, we present three types of inference strategies for the sample complier average causal effect: the Wald estimator, regression adjustment estimators and model-based Bayesian inference.  We derive asymptotic properties of the Wald estimator and regression adjustment estimators, under both complete randomization and Mahalanobis distance based rerandomization.  We use Monte Carlo simulation to compare the finite sample performance of different methods under both experimental designs.  

The next section defines the sample CACE.  Section 3 discusses complete randomization and Mahalanobis distance based rerandomization. Section 4 discusses the Wald estimator and regression adjustment estimators. Section 5 presents the model-based Bayesian inference. The small sample performance of
the methods are studied using Monte Carlo simulations in Section 6.  Section 7 presents an application to a job training experiment with non-compliance. The paper concludes with a discussion in Section 8.

\section{Sample Complier Average Causal Effect}

Consider an experiment with $n$ units. Let $\bm{x}_i$, $i=1,\cdots,n$, be the $K\times 1$ covariate vector for unit $i$. Let $\bm{X}$ be the $n\times K$ matrix of covariates for the $n$ units. Let $Z_i\in\{0,1\}$, $i=1,\cdots,n$, denote indicator of treatment assigned for unit $i$.  Let $\bm{Z}$ denote the vector of $Z_i$ for the $n$ units.  Suppose that there are $n_1$ units assigned to treatment and $n_0$ units assigned to control.  Not everyone assigned treatment/control receives treatment/control, however.  Due to this, one cannot estimate the average treatment effect without strong assumptions. It is however possible to estimate the average treatment effect on the compliers who would comply with treatment assignment.

We assume Stable Unit Treatment Value Assumption (SUTVA) \citep{Rubin_1978}, that is, there is only one version of treatment or control and there is no interference between units. Let $W_{i}(1)$ and $W_{i}(0)$, $i=1,\cdots,n$, be the potential treatment received if assigned to treatment and control for unit $i$.  Let $\bm{W}(z)$, $z=0,1$, be the vector of $W_i(z)$ for the $n$ units.  For $i=1,\cdots,n$, also define the four potential outcomes $Y_{i}(z,w)$ for which the treatment assigned and treatment received are fixed at $z=0,1$ and $w=0,1.$ For each individual only two of these potential outcomes, $Y_{i}(0,W_{i}(0))$ or $Y_{i}(1,W_{i}(1))$, can possibly be observed.  We further assume $Y_{i}(z,w)=Y_{i}(z^{\prime },w)$ for all $z,z^{\prime }$ and all $w$, that is, potential outcomes only depend on treatment received.  Thus the potential outcome if receiving treatment and if receiving control equals $Y_{i}(1)=Y_{i}(0,1)=Y_{i}(1,1)$ and $Y_{i}(0)=Y_{i}(0,0)=Y_{i}(1,0),$ respectively.  Let $\bm{Y}(w)$, $w=0,1$, be the vector of $Y_i(w)$ for the $n$ units.

There are four possible latent types of units: always-takers (\textit{at}) who would always receive treatment regardless of whether being assigned to treatment or control, with $W_i(1)=W_i(0)=1$; compliers (\textit{co}) who would receive treatment when assigned to treatment and receive control when assigned to control, with $W_i(1)=1$ and $W_i(0)=0$, or equivalently $W_i(1)-W_i(0)=1$; defiers (\textit{de}) who would receive control when assigned to treatment and receive treatment when assigned to control, with $W_i(1)=0$ and $W_i(0)=1$, or equivalently $W_i(1)-W_i(0)=-1$; and never-takers (\textit{nt}) who would always receive control regardless of whether being assigned to treatment or control, with $W_i(1)=W_i(0)=0$.

If we regard the units as being randomly sampled from a superpopulation, we can define the population complier average causal effect (CACE) as
\begin{equation}
\tau_{CACE}=\text{E}(Y_{i}(1)-Y_{i}(0)|W_{i}(1)-W_{i}(0)=1).
\label{pop_CACE}
\end{equation}%
The fraction of compliers in the superpopulation is $p_{co}=\text{Pr}(W_i(1)-W_i(0)=1)$.

\cite{Angrist_etal_1996} show that under further assumptions: (i) treatment assignment $Z_i$ being completely random, (ii) $P(W_i(1)=1)\neq P(W_i(0)=1)$ and (iii) $W_{i}(1)\geq W_{i}(0)$, it is possible to non-parametrically identify $\tau_{CACE}$. Assumption (ii) says that treatment assigned affects treatment received, and hence there exist compliers or defiers. Assumption (iii) rules out defiers.  Under assumption (iii), $p_{co}=\text{E}(W_i(1)-W_i(0))$.

In this paper we do not assume that the units in the study are randomly sampled from a superpopulation and instead focus only on the finite population consisting of these units.  This is of interest, for instance, when the units are volunteers who are different from the general population of interest.  We treat $\bm{W}(0)$, $\bm{W}(1)$, $\bm{Y}(0)$, $\bm{Y}(1)$ and $\bm{X}$ all as fixed, and treat only $\bm{Z}$ as random, unless stated explicitly otherwise.  Let $G_i,\ i=1,\cdots,n$ denote the latent group for unit $i$.  Since $W_i(1)$ and $W_i(0)$ are fixed, $G_i$ is also fixed.  The sample complier average causal effect (sample CACE) is defined as
\begin{equation}
\tau _{CACE}^{samp}=\frac{1}{n_{co}}\sum_{G_i=co}(Y_{i}(1)-Y_{i}(0)),
\label{samp_CACE}
\end{equation}
where $n_{co}=\sum_{i=1}^{n}I(W_{i}(1)-W_{i}(0)=1)$, with $I(\cdot)$ being the indicator function. $\tau _{CACE}^{samp}$ is a parameter for the finite population of units in experiment, for which point and interval estimates can be constructed.  The fraction of compliers in the sample is $p_{co}^{samp}=n_{co}/n$.

We adopt assumptions (ii) and (iii).  Under assumption (iii), there are no defiers, and $n_{co}=\sum_{i=1}^n (W_i(1)-W_i(0))$. Further under the assumption that potential outcomes only depend on treatment received, for always-takers and never-takers $Y_i(1,W_i(1))-Y_i(0,W_i(0))=0$, and for compliers $Y_i(1,W_i(1))-Y_i(0,W_i(0))=Y_i(1)-Y_i(0)$.  Therefore, $\sum_{G_i=co}(Y_{i}(1)-Y_{i}(0))=\sum_{i=1}^n (Y_i(1,W_i(1))-Y_i(0,W_i(0)))$.
Define the sample ITT effect on $W$ as
\begin{equation}
ITT_{W}^{samp}=\frac{1}{n}\sum_{i=1}^{n}(W_i(1)-W_i(0))=p_{co}^{samp},
\label{samp_ITT_W}
\end{equation}
and the sample ITT effect on $Y$ as
\begin{equation}
ITT_{Y}^{samp}=\frac{1}{n}\sum_{i=1}^{n}(Y_i(1,W_i(1))-Y_i(0, W_i(0))).
\label{sample_ITT_Y}
\end{equation}
We can then get another form of the sample CACE:
\begin{equation}
\tau _{CACE}^{samp}=\frac{ITT_{Y}^{samp}}{ITT_{W}^{samp}}.
\label{samp_CACE2}
\end{equation}

\section{Complete Randomization and Mahalanobis-distance Based Rerandomization}

We consider in this paper two experimental designs: complete randomization (CRE) and Mahalanobis-distance based rerandomization (ReM).

Under CRE, $n_1$ out of $n$ units are randomly assigned to treatment and the remaining $n_0$ units are assigned to control. There are $\tbinom{n}{n_{1}}=n_{A}$ possible treatment assignment  vectors.  Let $\bm{Z}^{j}$, $j=1,...,n_{A}$, denote the $j$th assignment vector, and $\mathbb{Z}=(\bm{Z}^{1},...,\bm{Z}^{n_{A}})$ the complete set of assignment vectors.

Let $\bm{x}_i$, $i=1,\cdots,n$, be the $K\times 1$ covariate vector for unit $i$. Let $\bm{X}$ be the $n\times K$ matrix of covariates for the $n$ units. Let $\bm{S}_{\bm{x}\bm{x}}$ denote the finite population covariance matrix of the covariates, which is defined
\begin{equation}
\bm{S}_{\bm{x}\bm{x}}=\frac{1}{n-1}\sum_{i=1}^{n}(\bm{x}_{i}-\bar{\bm{x}})(%
\bm{x}_{i}-\bar{\bm{x}})^{\top },  \label{eq:Cov}
\end{equation}
\noindent where $\bar{\bm{x}}=\sum_{i=1}^{n}\bm{x}_{i}/n$.

The Mahalanobis distance for the $j$th assignment vector is
\begin{equation}
M(\bm{Z}^{j},\bm{X)}=\frac{n}{4}(\widehat{\tau}_{\bm{x}}^{j\top}\bm{S}_{\bm{x}\bm{x}}^{-1}\widehat{\tau}_{\bm{x}}^{j}),\text{ }%
j=1,...,n_{A}  \label{eq:MD}
\end{equation}%
\noindent where
\begin{equation}
\widehat{\tau}_{\bm{x}}^{j}=\frac{1}{n_{1}}\sum_{i=1}^{n_{1}}Z_{i}^{j}%
\bm{x}_{i}-\frac{1}{n_{0}}\sum_{i=1}^{n_{0}}(1-Z_{i}^{j})\bm{x}_{i}=%
\overline{\bm{x}}_{1}^{j}-\overline{\bm{x}}_{0}^{j},\text{\ }%
j=1,...,n_{A},
\end{equation}
with $\overline{\bm{x}}_z^{j}$ being the mean value of $\bm{x}_i$ for those assigned to treatment arm $z$ ($z=0,1$).  \cite{morgan2012rerandomization} suggested accepting the treatment assignment vector $
\bm{Z}^{j}$ only when
\begin{equation}
M(\bm{Z}^{j},\bm{X)}\leq a,
\end{equation}%
where $a$ is a positive constant. This means that final randomization occur
only within the set
\begin{equation}
\mathcal{A}_{a}(\bm{X})=\{ \bm{Z}^{j}|M(\bm{Z}^{j},\bm{X)}%
\leq a\}.
\end{equation}
Asymptotically, the Mahalanobis distance follows a $\chi_K^2$ distribution (a chi-square distribution with $K$ degrees of freedom).  This implies that $a$ can be indirectly determined by setting $p_a=Pr(\chi_K^2\leq a)$.  For example, by setting $p_a=0.01$, $a$ equals to the 0.01 quantile of a $\chi_K^2$ distribution.

Compared to CRE, ReM better balances the mean values of covariates between the treated and control groups.

\section{Inference about Sample CACE using The Wald Estimator and Regression Adjustment Estimators}
\subsection{The Wald Estimator}

\subsubsection{Definition of The Wald Estimator}
Let the observed indicator of treatment received be $W_{i}^{obs}=W_{i}(Z_{i})$, $i=1,...,n$.  Let $\bm{W}^{obs}$ be the vector of $W_i^{obs}$ for all units.  Let the observed outcome be
$Y_{i}^{{obs}}=Y_{i}\left(W_{i}^{{obs}}\right)=Y_{i}\left(Z_i,W_{i}\left(Z_{i}\right)\right)$. Let $\bm{Y}^{obs}$ be the vector of $Y_i^{obs}$ for all units.
Define the estimator of ITT effect on treatment received as
\begin{equation}
\widehat{ITT}_{W}=\overline{W}_{1}^{{obs}}-\overline{W}_{0}^{{obs}%
},  \label{MDEW}
\end{equation}%
where
\begin{equation}
\overline{W}_{1}^{{obs}}=\frac{1}{n_{1}}\sum_{i:Z_{i}=1}W_{i}^{%
{obs}}\text{ and }\overline{W}_{0}^{{obs}}=\frac{1}{n_{0}}%
\sum_{i:Z_{i}=0}W_{i}^{{obs}}
\end{equation}
are the fractions of receiving treatment among those assigned to treatment and control.
Define the estimator of ITT effect on outcome as
\begin{equation}
\widehat{ITT}_{Y}=\overline{Y}_{1}^{{obs}}-\overline{Y}_{0}^{{obs}},  \label{ITTy}
\end{equation}%
where
\begin{equation}
\overline{Y}_{1}^{{obs}}=\frac{1}{n_{1}}\sum_{i:\ Z_i=1}Y_{i}^{{obs}}\text{ and }\overline{Y}_{0}=\frac{1}{n_{0}}\sum_{i:\ Z_i=0}Y_{i}^{{obs}}
\end{equation}
are the mean observed outcomes among those assigned to treatment and control.

The Wald estimator is defined as
\begin{equation}
\widehat{\tau}_{CACE}^{Wald}=\frac{\widehat{ITT}_{Y}}{\widehat{ITT}_{W}}=\frac{\overline{Y}_{1}^{obs}-\overline{Y}_{0}^{obs}}{%
\overline{W}_{1}^{obs}-\overline{W}_{0}^{obs}}.  \label{est_Wald}
\end{equation}
In the case when the units are regarded as being randomly sampled from a superpopulation and treatment assignment is completely random, $\widehat{\tau}_{CACE}^{Wald}$ is a asymptotically unbiased and normally distributed estimator for $\tau_{CACE}$, and the delta method can be used to estimate its large-sample sampling variance.

We now study the properties of using $\widehat{\tau}_{CACE}^{Wald}$ to estimate $\tau_{CACE}^{samp}$.

\subsubsection{Properties of the Wald Estimator under Complete Randomization}
For $z=0,1$, denote the finite population variance of $Y_i(z, W_i(z))$ as
\begin{equation}
	\mathbb{S}_{Y_z}^2=\frac{1}{n-1}\sum_{i=1}^{n}\left(Y_i(z, W_i(z))-\overline{Y}_z\right)^2\text{ for }\overline{Y}_z=\frac{1}{n}\sum_{i=1}^{n}Y_i(z, W_i(z)).
\end{equation}
The following theorem says that $\widehat{\tau}_{CACE}^{Wald}$ is a consistent estimator of $\tau_{CACE}^{samp}$ under certain conditions.

\setcounter{theorem}{0}
\begin{theorem}
	\label{consistency_CRE}
	Assume that the limit inferior of $p_{co}^{samp}$ is positive and that the limit superiors of $ITT^{samp}_Y$, $\mathbb{S}_{Y_1}^2$ and $\mathbb{S}_{Y_0}^2$ are finite. Also assume that the proportion of units under treatment $n_1/n$ has a limit in $(0,1)$. Under CRE,
$\widehat{\tau}_{CACE}^{Wald}-\tau_{CACE}^{samp}=o_p(1)$ as $n\rightarrow\infty$.
\end{theorem}

We next discuss construction of confidence interval for $\tau_{CACE}^{samp}$.  Define
\begin{equation}
 A_i(z)=Y_i(z,W_i(z))-W_i(z)\tau_{CACE}^{samp},\ \ z=0,1.\label{Az}
 \end{equation} Define
	\begin{equation}
		\widetilde{\tau}_A=\frac{1}{n_1}\sum_{i:\ Z_i=1}A_i(1)-\frac{1}{n_0}\sum_{i:\ Z_i=0}A_i(0)=\widehat{ITT}_{Y}-\widehat{ITT}_{W}\tau_{CACE}^{samp},\label{tauA_tilde}
	\end{equation}
the difference between the mean value of $A_i(1)$ for those assigned treatment and the mean value of $A_i(0)$ for those assigned control.
Applying Theorem 3 in \cite{li2017general}, over all randomizations, $\widetilde{\tau}_A$ has mean
\begin{equation}
\tau_A = \frac{1}{n}\sum_{i=1}^n (A_i(1) -A_i(0))=ITT_{Y}^{samp}-ITT_{W}^{samp}\tau_{CACE}^{samp}=0
\end{equation}
 due to \eqref{samp_CACE2}, and variance \begin{equation}Var\left(\widetilde{\tau}_A\right)=\frac{\mathbb{S}_{A_1}^2}{n_1}+\frac{\mathbb{S}_{A_0}^2}{n_0}-\frac{\mathbb{S}_{A_{01}}^2}{n}.
\label{var_tauhat_A}
	\end{equation}
Here $\mathbb{S}_{A_z}^2$ is the finite population variance of $A_i(z)$ for $z=0,1$, that is,
\begin{equation}
\mathbb{S}_{A_z}^2=\frac{1}{n-1}\sum_{i=1}^n (A_i(z)-\bar{A}_z)^2,
\end{equation}
where $\bar{A}_z=1/n \sum_{i=1}^n A_i(z)$;
$\mathbb{S}_{A_{01}}^2$ is the finite population variance of of $A_i(1)-A_i(0)$, that is,
\begin{equation}
\mathbb{S}_{A_{01}}^2=\frac{1}{n-1}\sum_{i=1}^n (A_i(1)-A_i(0))^2.
\end{equation}
Furthermore, by applying Theorem 4 in \cite{li2017general} to two treatment groups and a single contrast, $\widetilde{\tau}_A/\sqrt{Var(\widetilde{\tau}_A)}\stackrel{d}{\longrightarrow}N(0,1)$ as $n\rightarrow\infty$ under the condition
	\begin{equation}
		\label{lindeberg-feller}
		\lim_{n\rightarrow\infty}\frac{\max_{1 \leqslant i \leqslant n}\left(A_{i}(z)-\overline{A}(z)\right)^2}{\min(n_1, n_0)^2\cdot Var\left(\widetilde{\tau}_A\right)}=0,\quad\text{for }z=0,1.
	\end{equation}
	
Let $\nu_{1-\alpha/2}$ be the $(1-\alpha/2)$th quantile of $N(0,1)$.  As $n\rightarrow\infty$, the interval
	\begin{equation}
		\widetilde{\tau}_A\pm \nu_{1-\alpha/2}\sqrt{Var(\widetilde{\tau}_A)}
	\end{equation}
	has a $(1-\alpha)$ coverage rate for $\tau_A=0$.  Combining this with \eqref{tauA_tilde} and \eqref{est_Wald}, it is easy to show that an equivalent statement is that, as $n\rightarrow\infty$,
	\begin{equation}
		\label{ci_CRE1}
		\widehat{\tau }_{CACE}^{Wald}\pm \nu_{1-\alpha/2}\sqrt{Var(\widetilde{\tau}_A)}/\widehat{ITT}_{W}
	\end{equation}
	has a $(1-\alpha)$ coverage rate for $\tau_{CACE}^{samp}$.

If we can find an asymptotically conservative estimator for $Var\left(\widetilde{\tau}_A\right)$, we can construct an asymptotically conservative $(1-\alpha)$ confidence interval for $\tau_{CACE}^{samp}$.  Replacing $\tau_{CACE}^{samp}$ in the expression of $A_i(z)$ in \eqref{Az} with its estimator $\widehat{\tau}_{CACE}^{Wald}$, we define
\begin{equation}
\widehat{A}_i(z)=Y_{i}(z,W_i(z))-W_i(z)\widehat{\tau}_{CACE}^{Wald}.\label{Azhat}
\end{equation}  We can use the sample variance of $\widehat{A}_i(z)$ under treatment arm $z$,
\begin{equation}
S_{\widehat{A}(z)}^2=\frac{1}{n_z-1}\sum_{i:Z_i=z}\left(\widehat{A}_i(z)-\sum_{i:Z_i=z}\widehat{A}_i(z)/n_z\right)^2,
\end{equation}
to estimate $\mathbb{S}_{A_z}^2$ for $z=0,1$.  The proposed variance estimator is
	\begin{equation}		\widehat{Var}(\widetilde{\tau}_A)=\frac{S_{\widehat{A}(1)}^2}{n_1}+\frac{S_{\widehat{A}(0)}^2}{n_0}.
	\end{equation}
The following proposition shows that $\widehat{Var}(\widetilde{\tau}_A)$ is asymptotically equivalent to
	\begin{equation}
		Var\left(\widetilde{\tau}_A\right)^{+}=\frac{\mathbb{S}_{A_1}^2}{n_1}+\frac{\mathbb{S}_{A_0}^2}{n_0},
	\end{equation}
and hence is asymptotically conservative for $Var\left(\widetilde{\tau}_A\right)$.

\setcounter{theorem}{0}
	\begin{proposition}
		\label{conservative_CRE}
		Assume that \eqref{lindeberg-feller} holds, and that (i) the limit inferior of $p_{co}^{samp}$ is positive; (ii) the limit of the proportion of units under treatment is in $(0,1)$.  Under CRE, $\widehat{Var}(\widetilde{\tau}_A)/Var(\widetilde{\tau}_A)^{+}$
$\stackrel{p}{\longrightarrow}1$ as $n\rightarrow\infty$.
	\end{proposition}

An asymptotically conservative $(1-\alpha)$ confidence interval for $\tau_{CACE}^{samp}$ is then given by
	\begin{equation}
		\label{ci_Wald}
		\widehat{\tau }_{CACE}^{Wald}\pm \nu_{1-\alpha/2}\sqrt{\widehat{Var}(\widetilde{\tau}_A)}/\widehat{ITT}_{W}.
	\end{equation}
Proposition A1 in the Appendix shows that this confidence interval is the same as the super-population confidence interval for $\tau_{CACE}$ obtained by the delta method (see e.g. \cite{imbens2015causal}, Chapter 23).
	
\subsubsection{Properties of the Wald Estimator under Rerandomization}

We need the following lemma to show that $\tau_{CACE}^{Wald}$ is a consistent estimator of $\tau_{CACE}^{samp}$ under ReM.

\setcounter{theorem}{0}
\begin{lemma}
	\label{dominated_converge_in_p}
For any real number $k$, if a sequence of random variables $\{U_n\}_{n=1}^{\infty}$ is $o_p(n^k)$ under CRE as $n\rightarrow\infty$, it is also $o_p(n^k)$ under ReM as $n\rightarrow\infty$.
\end{lemma}

When conditions in Theorem 1 are satisfied, by applying Theorem 1 and Lemma 1, we immediately obtain that $\widehat{\tau}^{Wald}_{CACE}-\tau^{samp}_{CACE}=o_p(1)$ under ReM as $n\rightarrow\infty$.

We next show that an asymptotically conservative confidence interval for $\tau_{CACE}^{samp}$ can be constructed under ReM, and that it is narrower than the confidence interval constructed under CRE.

According to \cite{li2018asymptotic}, under ReM, under certain conditions,
\begin{equation}
\widetilde{\tau}_A/\sqrt{Var(\widetilde{\tau}_A)}\overset{d}{\rightarrow }\sqrt{1-R^2} \cdot \varepsilon_0+\sqrt{R^2} \cdot L_{K, a}.\label{mix_normal_trunc}
\end{equation}
Here $\varepsilon _{0}$ is a standard normal variable, which is related
to the space orthogonal to that of the covariates and hence is unaffected by
rerandomization, $L_{K,a}$ is related to the linear projection of $A_i(z)$
($z=0,1$) into the space of covariates and is thus affected by
rerandomization, and
\begin{equation}
	R^2=\left(\frac{\mathbb{S}^2_{A_1 \mid \boldsymbol{x}}}{n_1}+\frac{\mathbb{S}^2_{A_0 \mid \boldsymbol{x}}}{n_0}-\frac{\mathbb{S}^2_{A_{01}\mid \boldsymbol{x}}}{n}\right)/Var(\widetilde{\tau}_A),
\end{equation}
where $S_{A_z|\bm{x}}$ ($z=0,1$) and $S_{A_{01}|\bm{x}}$ denote, respectively, the finite population
variances of the linear projection of $A_i(z)$ and $A_i(1)-A_i(0)$ on $\bm{x}_i$.

The distribution of $L_{K,a}$ has the following form%
\begin{equation}
L_{K,a}\sim \chi_{K,a}F\sqrt{\beta _{K}},
\end{equation}%
where $\chi_{K,a}=\sqrt{\chi_{K,a}^2}$, with $\chi_{K,a}^2=\chi_{K}^{2}|\chi_{K}^{2}\leq a$ being a truncated $\chi^{2}$ random variable, $F$ is a random variable taking values $\pm 1$ with
probability 1/2, $\beta_{K}\sim \text{Beta}(1/2,(K-1)/2)$ is a Beta random
variable degenerating to a point mass at 1 when $K=1$, and ($\chi _{K,a}$,$F$, $\beta_{K}$) are jointly independent. The distribution of $L_{K,a}$ is symmetric around zero, and is more concentrated around zero than the normal distribution.

For any given value of $R^2$, let $\lambda_{1-\alpha/2, a}(R^2)$ denote the $1-\alpha/2$th quantile of the distribution of $\sqrt{1-R^2} \cdot \varepsilon_0+\sqrt{R^2} \cdot L_{K, a}$ in \eqref{mix_normal_trunc}. As $n\rightarrow\infty$, the interval
\begin{equation}
	\widetilde{\tau}_A\pm\lambda_{1-\alpha/2,a}(R^2)\sqrt{Var}(\widetilde{\tau}_A)
\end{equation}
has an $(1-\alpha)$ coverage rate for $\tau_A=0$. Combining this with \eqref{tauA_tilde} and \eqref{est_Wald}, it is easy to show that an equivalent statement is that, as $n\rightarrow\infty$,
\begin{equation}
	\label{ci_ReM}	\widehat{\tau}_{CACE}^{Wald}\pm\lambda_{1-\alpha/2,a}(R^2)\sqrt{Var}(\widetilde{\tau}_A)/\widehat{ITT}_{W}
\end{equation}
has an $(1-\alpha)$ coverage rate for $\tau_{CACE}^{samp}$.

Since the distribution of $L_{K,a}$ is more concentrated around zero than the standard normal distribution, as $R^2$ decreases, $\lambda_{1-\alpha/2, a}(R^2)$ increases.  Therefore, with an estimator of $R^2$ whose asymptotic bias is zero or negative and an asymptotically conservative estimator for $Var(\widetilde{\tau}_A)$, we can construct an asymptotically conservative $(1-\alpha)$ confidence interval for $\tau_{CACE}^{samp}$.

Let $\mathbb{S}_{A_z,\boldsymbol{x}}$ be the finite population covariance between $A_i(z)$ and $\boldsymbol{x}_i$,
\begin{equation} \mathbb{S}_{A_z,\boldsymbol{x}}=\frac{1}{n-1}\sum_{i=1}^{n}(A_i(z)-\overline{A}_z)(\bm{x}_i-\overline{\bm{x}}).
\end{equation}
The finite population variances of the linear projections of $A_i(z)$ and $A_i(1)-A_i(0)$ on $\bm{x}_i$ are:
	\begin{align}
		&\mathbb{S}^2_{A_z| \boldsymbol{x}}=\mathbb{S}_{A_z,\boldsymbol{x}}\mathbb{S}_{\bm{x}\bm{x}}^{-1}\mathbb{S}_{A_z,\boldsymbol{x}}^{\top},\ z=0,1\label{projAz}\\
		&\mathbb{S}^2_{A_{01}\mid \boldsymbol{x}}=\left(\mathbb{S}_{A_1, \boldsymbol{x}}-\mathbb{S}_{A_0, \boldsymbol{x}}\right)\mathbb{S}_{\bm{x}\bm{x}}^{-1}\left(\mathbb{S}_{A_1,\boldsymbol{x}}^{\top}-\mathbb{S}_{A_0,\boldsymbol{x}}^{\top}\right).\label{projA01}
	\end{align}

Recalling that $\widehat{A}_i(z)=Y_{i}(z,W_i(z))-W_i(z)\widehat{\tau }_{CACE}^{Wald}$, we can use the sample covariance between $\widehat{A}_i(z)$ and $\boldsymbol{x}_i$ under treatment arm $z$,
\begin{equation*} S_{\widehat{A}_z,\boldsymbol{x}}=\frac{1}{n_z-1}\sum_{i:Z_i=z}\left(\widehat{A}_i(Z_i)-\sum_{i:Z_i=z}\widehat{A}_i(Z_i)/n_z\right)\left(\boldsymbol{x}_i-\sum_{i:Z_i=z}\boldsymbol{x}_i/n_z\right),
\end{equation*}
to estimate $\mathbb{S}_{A_z,\boldsymbol{x}}$ for $z=0,1$.  Replacing $\mathbb{S}_{A_z,\boldsymbol{x}}$ in the expressions in \eqref{projAz} and \eqref{projA01} with $S_{\widehat{A}(z),\boldsymbol{x}}$, we obtain $S^2_{\widehat{A}_z| \boldsymbol{x}}$ and $S^2_{\widehat{A}_{01}| \boldsymbol{x}}$ as estimators of $\mathbb{S}^2_{A_z| \boldsymbol{x}}$ and $\mathbb{S}^2_{A_{01}| \boldsymbol{x}}$.

We propose the following estimator for $Var(\widetilde{\tau}_A)$:
\begin{equation}	\widehat{Var}(\widetilde{\tau}_A)_{\bm{x}}=\frac{S_{\widehat{A}_1}^2}{n_1}+\frac{S_{\widehat{A}_0}^2}{n_0}-\frac{S^2_{\widehat{A}_{01}\mid \boldsymbol{x}}}{n},
\end{equation}
and the following estimator for $R^2$:
\begin{equation}
	\widehat{R}^2=\left(\frac{S_{\widehat{A}_1\mid \boldsymbol{x}}^2}{n_1}+\frac{S_{\widehat{A}_0\mid \boldsymbol{x}}^2}{n_0}-\frac{S^2_{\widehat{A}_{01}\mid \boldsymbol{x}}}{n}\right)/\widehat{Var}(\widetilde{\tau}_A)_{\bm{x}}.
\end{equation}

In the next proposition, we show that $\widehat{Var}(\widetilde{\tau}_A)_{\bm{x}}$ and $\widehat{R}^2$ are asymptotically equivalent to
\begin{equation} Var\left(\widetilde{\tau}_A\right)_{\bm{x}}^+=\frac{\mathbb{S}_{A_1}^2}{n_1}+\frac{\mathbb{S}_{A_0}^2}{n_0}-\frac{\mathbb{S}^2_{A_{01}\mid \boldsymbol{x}}}{n}
\end{equation}
and
\begin{equation}	R^{2-}=\left(\frac{\mathbb{S}_{A_1\mid\boldsymbol{x}}^2}{n_1}+\frac{\mathbb{S}_{A_0\mid\boldsymbol{x}}^2}{n_0}-\frac{\mathbb{S}^2_{A_{01}\mid\boldsymbol{x}}}{n}\right)/Var\left(\widetilde{\tau}_A\right)_{\bm{x}}^+,
\end{equation}
respectively.  Since $\mathbb{S}^2_{A_{01}\mid \boldsymbol{x}}\leqslant\mathbb{S}^2_{A_{01}}$, we have $Var\left(\widetilde{\tau}_A\right)_{\bm{x}}^+\geqslant Var\left(\widetilde{\tau}_A\right)$ and $R^{2-}\leqslant R^{2}$.  Hence $\widehat{Var}(\widetilde{\tau}_A)_{\bm{x}}$ is a conservative estimator of $Var\left(\widetilde{\tau}_A\right)$, and the asymptotic bias of $\widehat{R}^2$ is zero or negative.

\setcounter{theorem}{0}
\begin{condition}
	\label{strict_condition}
	As $n\rightarrow\infty$, for $z=0,1$, (i) the limit of the proportion of units under treatment is in $(0,1)$; (ii) the finite population variances and covariances $\mathbb{S}_{A_z}^2$, $\mathbb{S}_{A_{01}}^2$, $\mathbb{S}_{\boldsymbol{x}\boldsymbol{x}}$ and $\mathbb{S}_{A_z,\boldsymbol{x}}$ have finite limiting values for $z=0,1$, and the limit of $\mathbb{S}_{\boldsymbol{x}\boldsymbol{x}}$ is non-singular; (iii) $\max_{1\leqslant i\leqslant n}\left\|A_i(z)-\overline{A}_z\right\|_{2}^{2} / n \rightarrow 0$ and $\max_{1\leqslant i\leqslant n}\left\|\boldsymbol{x}_{i}-\overline{\boldsymbol{x}}\right\|_{2}^{2}/n$ $\rightarrow 0$.
\end{condition}

\begin{proposition}
	\label{conservative_rem}
	Assume that Condition \ref{strict_condition} holds and that the limit inferior of $p_{co}^{samp}$ is positive.  Under ReM, $\widehat{Var}(\widetilde{\tau}_A)_{\bm{x}}-Var\left(\widetilde{\tau}_A\right)_{\bm{x}}^+=o_p(n^{-1})$ and $\widehat{R}^2-R^{2-}=o_p(1)$ as $n\rightarrow\infty$.
\end{proposition}

We can now construct an asymptotically conservative $1-\alpha$ confidence interval for $\tau_{CACE}^{samp}$ as
\begin{equation}
	\label{ci_ReM}
	\widehat{\tau}_{CACE}^{Wald}\pm \nu_{1-\alpha/2,a}(\widehat{R}^2)\sqrt{\widehat{Var}(\widetilde{\tau}_A)_{\bm{x}}}/\widehat{ITT}_{W}.
\end{equation}

\subsection{Regression Adjustment Estimators}

We centralize the covariates by letting $\bm{x}_i^*=\bm{x}_i-\bar{\bm{x}}$, where $\bm{x}=\sum_{i=1}^n \bm{x}_i/n$.  Under CRE, according to the results of \cite{lin2013agnostic}, $ITT_W^{samp}$ or $ITT_Y^{samp}$ can be consistently estimated by the estimated coefficient on $Z_i$ (denoted by $\widehat{ITT}_W^{adj}$ or $\widehat{ITT}_Y^{adj}$) in the Ordinary Least squares (OLS) regression of $W_i^{obs}$ on 1, $Z_i$, $\bm{x}_i^*$ and $Z_i\bm{x}_i^*$ using all units, and asymptotically valid confidence intervals for $ITT_W^{samp}$ or $ITT_Y^{samp}$ can be constructed using the Eicker-Huber-White (EHW) robust standard error estimator.  According to \cite{li2020rerandomization}, similar results hold under ReM.

For $z=0,1$, let $\widehat{\bm{\beta}}_{W,z}$ be the vector of estimated coefficients on $\bm{x}_i^*$ in the OLS regression of $W_i^{obs}$ on $\bm{x}_i^*$ using units with $Z_i=z$, and let $\widehat{\bm{\beta}}_{Y,z}$ be the vector of estimated coefficients on $\bm{x}_i^*$ in the OLS regression of $Y_i^{obs}$ on 1 and $\bm{x}_i^*$ using units with $Z_i=z$.  Applying Proposition A2 in the appendix, we have
	\begin{equation}
\begin{aligned}
		\widehat{ITT}_W^{adj}&=\frac{1}{n_1}\sum_{i:\ Z_i=1}\left(W_i^{obs}-\widehat{\bm{\beta}}_{W,1}\bm{x}_i^*\right)-\frac{1}{n_0}\sum_{i:\ Z_i=0}\left(W_i^{obs}-\widehat{\bm{\beta}}_{W,0}\bm{x}_i^*\right)\\
&=\widehat{ITT}_W - \widehat{\bm{\beta}}_{W,1}\overline{\bm{x}}_1^*+\widehat{\bm{\beta}}_{W,0}\overline{\bm{x}}_0^*,
\end{aligned}\label{ITThat_W_adj}
	\end{equation}
	\begin{equation}
\begin{aligned}
		\widehat{ITT}_Y^{adj}&=\frac{1}{n_1}\sum_{i:\ Z_i=1}\left(Y_i^{obs}-\widehat{\bm{\beta}}_{Y,1}\bm{x}_i^*\right)-\frac{1}{n_0}\sum_{i:\ Z_i=0}\left(Y_i^{obs}-\widehat{\bm{\beta}}_{Y,0}\bm{x}_i^*\right)\\
&=\widehat{ITT}_Y - \widehat{\bm{\beta}}_{Y,1}\overline{\bm{x}}_1^*+\widehat{\bm{\beta}}_{Y,0}\overline{\bm{x}}_0^*.
\end{aligned}\label{ITThat_Y_adj}
	\end{equation}
Define the regression adjustment estimator of $\tau_{CACE}^{samp}$ as
\begin{equation}
\widehat{\tau}_{CACE}^{adj}=\frac{\widehat{ITT}_{Y}^{adj}}{\widehat{ITT}_{W}^{adj}}.  \label{est_adj}
\end{equation}

As CRE can be viewed as a special case of ReM with $a=\infty$, in this section we only need to state theoretical results for ReM.  The following theorem says that $\widehat{\tau}_{CACE}^{adj}$ is a consistent estimator of $\tau_{CACE}^{samp}$.

\setcounter{theorem}{1}
	\begin{theorem}
	\label{consistency_ReM_adj}
	Assume that Condition \ref{strict_condition} holds, that the limit inferior of $p_{co}^{samp}$ is positive, and that the limit superiors of $ITT^{samp}_Y$ is finite. Under ReM, $\widehat{\tau}_{CACE}^{adj}-\tau_{CACE}^{samp}=o_p(1)$ as $n\rightarrow\infty$.
	\end{theorem}

We next discuss construction of confidence interval for $\tau_{CACE}^{samp}$.  For $z=0,1$, let $\bm{\beta}_{W,z}$ be the vector of coefficients on $\bm{x}_i^*$ in the linear projection of $W_i(z)$ on $\bm{x}_i^*$ using all units, and let $\bm{\beta}_{Y,z}$ be the vector of coefficients on $\bm{x}_i^*$ in the linear projection of $Y_i(z,W_i(z))$ on $\bm{x}_i^*$ using all units.
Define
\begin{equation}
B_i(z)=\left(Y_{i}(z,W_i(z))-\bm{\beta}_{Y,z}\bm{x}_i^*\right)-\left(W_i(z)-\bm{\beta}_{W,z}\bm{x}_i^*\right)\tau_{CACE}^{\text{samp}},\ \ z=0,1.\label{Bz}
\end{equation}
Recall that $A_i(z)=Y_{i}(z,W_i(z))-W_i(z)\tau_{CACE}^{samp}$, which is a linear combination of $Y_i(z,W_i(z))$ and $W_i(z)$.  Therefore, the vector of coefficients on $\bm{x}_i^*$ in the linear projection of $A_i(z)$ on $\bm{x}_i^*$ using all units is $\bm{\beta}_{A,z}=\bm{\beta}_{Y,z}-\bm{\beta}_{W,z}\tau_{CACE}^{samp}$, and we can write $B_i(z)=A_i(z)-\bm{\beta}_{A,z}\bm{x}_i^*$.
Define
	\begin{equation}
		\widetilde{\tau}_B=\frac{1}{n_1}\sum_{i:\ Z_i=1}B_i(1)-\frac{1}{n_0}\sum_{i:\ Z_i=0}B_i(0),\label{tauB_tilde}
	\end{equation}
the difference in mean value of $B_i(1)$ for those assigned treatment and the mean value of $B_i(0)$ for those assigned control.  According to Theorem 3 in \cite{li2020rerandomization}, under Condition  \ref{strict_condition} and ReM,
\begin{equation}
(\widetilde{\tau}_B-\tau_A)/\sqrt{Var(\widetilde{\tau}_B)}=\widetilde{\tau}_B/\sqrt{Var(\widetilde{\tau}_B)}\stackrel{d}{\longrightarrow}N(0,1),
\end{equation}
where $Var(\widetilde{\tau}_B)=(1-R^2)Var(\widetilde{\tau}_A)$.

For $z=0,1$, let $\bm{\beta}_{A,z}'$ be the vector of coefficients on $\bm{x}_i^*$ in the linear projection of $A_i(z)$ on $\bm{x}_i^*$ using units with $Z_i=z$.  Again because $A_i(z)$ is a linear combination of $Y_i(z,W_i(z))$ and $W_i(z)$, we have $\bm{\beta}_{A,z}'=\widehat{\bm{\beta}}_{Y,z}-\widehat{\bm{\beta}}_{W,z}\tau_{CACE}^{samp}$.  Define
\begin{equation}
B_i'(z)=A_i(z)-\bm{\beta}_{A,z}'\bm{x}_i^*,\label{Bz_alt}
\end{equation}
and
	\begin{equation}
		\widetilde{\tau}_B'=\frac{1}{n_1}\sum_{i:\ Z_i=1}B_i'(1)-\frac{1}{n_0}\sum_{i:\ Z_i=0}B_i'(0).\label{tauB_tilde_alt}
	\end{equation}
Plugging in the definition of $B_i'(z)$ into \eqref{tauB_tilde_alt} and combining with \eqref{ITThat_W_adj}
and \eqref{ITThat_Y_adj}, it is easy to show that
\begin{equation}
\widetilde{\tau}_B'=\widehat{ITT}_Y^{adj}-\widehat{ITT}_W^{adj}\tau_{CACE}^{samp}.\label{tauB_tilde_alt2}
\end{equation}
According to Proposition 3 in \cite{li2020rerandomization}, under Condition  \ref{strict_condition} and ReM,
$\widetilde{\tau}_B'$ has the same asymptotic distribution as $\widetilde{\tau}_B$, and hence
\begin{equation}
\widetilde{\tau}_B'/\sqrt{Var(\widetilde{\tau}_B')}\stackrel{d}{\longrightarrow}N(0,1),
\end{equation}
where $Var(\widetilde{\tau}_B')=(1-R^2)Var(\widetilde{\tau}_A)$.

Recall that $\nu_{1-\alpha/2}$ denotes the $(1-\alpha/2)$th quantile of $N(0,1)$.  As $n\rightarrow\infty$, the interval
	\begin{equation}
		\widetilde{\tau}_B'\pm \nu_{1-\alpha/2}\sqrt{Var(\widetilde{\tau}_B')}
	\end{equation}
	has a $(1-\alpha)$ coverage rate for 0.
Combining this with \eqref{tauB_tilde_alt2} and \eqref{est_adj}, it is easy to show that an equivalent statement is that, as $n\rightarrow\infty$,
	\begin{equation}
		\label{ci_CRE1}
		\widehat{\tau}_{CACE}^{adj}\pm \nu_{1-\alpha/2}\sqrt{Var(\widetilde{\tau}_B')}/\widehat{ITT}_{W}^{adj}
	\end{equation}
	has a $(1-\alpha)$ coverage rate for $\tau_{CACE}^{samp}$.

Applying Proposition A2 in the appendix, we know that $\widetilde{\tau}_B'$ equals the coefficient on $Z_i$ in the linear projection of $A_i(Z_i)$ on 1, $Z_i$, $\bm{x}_i^*$ and $Z_i\bm{x}_i^*$ using all units.  If $A_i(Z_i)$ was known, we could use the EHW robust variance estimator to estimate $Var(\widetilde{\tau}_B')$.  However, $A_i(Z_i)$ contains the unknown quantity $\tau_{CACE}^{samp}$.  Hence in practice we replace $\tau_{CACE}^{samp}$ with $\widehat{\tau}_{CACE}^{adj}$, and perform instead linear projection of
\begin{equation}
\widehat{A}_i^{adj}=Y_i(Z_i,W_i(Z_i))-W_i(Z_i)\widehat{\tau}_{CACE}^{adj}=Y_i^{obs}-W_i^{obs}\widehat{\tau}_{CACE}^{adj}.
\end{equation}
Let $\widehat{V}_{EHW}$ denote the EHW variance estimator of the coefficient on $Z_i$ in the linear projection of $\widehat{A}_i^{adj}$ on 1, $Z_i$, $\bm{x}_i^*$ and $Z_i\bm{x}_i^*$ using all units.

 The next proposition shows $\widehat{V}_{EHW}$ is asymptotically equivalent to $(1-R^{2-})\operatorname{Var}\left(\widetilde{\tau}_A\right)_{\bm{x}}^{+}$, and hence is a conservative estimator of $Var(\widetilde{\tau}_B')=(1-R^{2})Var(\widetilde{\tau}_A)$.
\begin{proposition}
	\label{conservative_reg}
	Under ReM and Condition \ref{strict_condition}, $\widehat{V}_{EHW}-(1-R^{2-})\operatorname{Var}\left(\widetilde{\tau}_A\right)_{\bm{x}}^{+}=o_p(n^{-1})$ as $n\rightarrow\infty$.
\end{proposition}
Thus, an asymptotically conservative $1-\alpha$ confidence interval for $\tau_{CACE}^{samp}$ can be constructed as
\begin{equation}
	\label{ci_Reg_HW}
	\widehat{\tau}_{CACE}^{adj}\pm \nu_{1-\alpha/2}\sqrt{\widehat{V}_{EHW}}/\widehat{ITT}_{W}^{adj}.
\end{equation}

We can also use the series of HC variance estimators that serve as finite sample corrections of the EHW variance estimator \citep{mackinnon2012thirty}. Specifically, let $\widehat{V}_{HC2}$ and $\widehat{V}_{HC3}$ denote the HC2 and HC3 variance estimators of the coefficient on $Z_i$ in the linear projection of $\widehat{A}_i^{adj}$ on 1, $Z_i$, $\bm{x}_i^*$ and $Z_i\bm{x}_i^*$ using all units.  Although $\widehat{V}_{HC2}$ and $\widehat{V}_{HC3}$ are more conservative than $\widehat{V}_{EHW}$ in finite samples, they are asymtotically equivalent as shown in the next proposition.
\begin{proposition}
	\label{conservative_hc}
	Under ReM and Condition \ref{strict_condition}, $\widehat{V}_{HCj}-(1-R^{2-})\operatorname{Var}\left(\widetilde{\tau}_A\right)_{\bm{x}}^{+}=o_p(n^{-1})$ as $n\rightarrow\infty$ for $j=2,3$.
\end{proposition}
Thus, another two asymptotically conservative $1-\alpha$ confidence intervals for $\tau_{CACE}^{samp}$ can be constructed as
\begin{equation}
	\label{ci_Reg_HC}
	\begin{aligned}
		&\widehat{\tau }_{CACE}^{adj}\pm \nu_{1-\alpha/2}\sqrt{\widehat{V}_{HC2}}/\widehat{ITT}_{W}^{adj};\\
		&\widehat{\tau }_{CACE}^{adj}\pm \nu_{1-\alpha/2}\sqrt{\widehat{V}_{HC3}}/\widehat{ITT}_{W}^{adj}.
	\end{aligned}
\end{equation}

\section{Bayesian Inference about Sample CACE}

\subsection{General Framework}
Let the missing indicator of treatment received be $W_{i}^{mis}=W_{i}(1-Z_{i})$, $i=1,...,n$.  Let $\bm{W}^{mis}$ be the vector of $W_i^{obs}$ for all units.  Let the missing outcome be
$Y_{i}^{{mis}}=Y_{i}\left(W_{i}^{{mis}}\right)=Y_{i}\left(1-Z_i,W_{i}\left(1-Z_{i}\right)\right)$.  Let $\bm{Y}^{mis}$ be the vector of $Y_i^{obs}$ for all units.

To estimate $\tau_{CACE}^{samp}$ using a model-based Bayesian approach, a superpopulation model is posed, treating $\bm{W}(0)$, $\bm{W}(1)$, $\bm{Y}(0)$ and $\bm{Y}(1)$ all as random.  $\bm{W}^{mis}$ and $\bm{Y}^{mis}$ are then imputed based on their joint posterior distribution conditional on $\bm{W}^{obs}$, $\bm{Y}^{obs}$, $\bm{Z}$ and $\bm{X}$:
\begin{equation}
\begin{aligned}
&Pr(\bm{W}^{mis},\bm{Y}^{mis}|\bm{W}^{obs},\bm{Y}^{obs},\bm{Z},\bm{X})\\
=&
\frac{Pr(\bm{W}^{mis},\bm{W}^{obs},\bm{Y}^{mis},\bm{Y}^{obs},\bm{Z}|\bm{X})}{Pr(\bm{W}^{obs},\bm{Y}^{obs},\bm{Z}|\bm{X})}\\
=&\frac{Pr(\bm{W}(0),\bm{W}(1),\bm{Y}(0),\bm{Y}(1)|\bm{X})Pr(\bm{Z}|\bm{W}(0),\bm{W}(1),\bm{Y}(0),\bm{Y}(1),\bm{X})}
{\int Pr(\bm{W}(0),\bm{W}(1),\bm{Y}(0),\bm{Y}(1)|\bm{X})Pr(\bm{Z}|\bm{W}(0),\bm{W}(1),\bm{Y}(0),\bm{Y}(1),\bm{X})d\bm{W}^{mis}d\bm{Y}^{mis}}.
\end{aligned}\label{Bayesian_eq1}
\end{equation}
It is important that under both CRE and ReM, treatment assignment does not depend on $\bm{W}(0)$, $\bm{W}(1)$, $\bm{Y}(0)$ and $\bm{Y}(1)$, we can write $Pr(\bm{Z}|\bm{W}(0),\bm{W}(1),\bm{Y}(0),\bm{Y}(1),\bm{X})=Pr(\bm{Z}|\bm{X})$, and simplify \eqref{Bayesian_eq1} to be
\begin{equation}
\begin{aligned}
&Pr(\bm{W}^{mis},\bm{Y}^{mis}|\bm{W}^{obs},\bm{Y}^{obs},\bm{Z},\bm{X})\\
=&\frac{Pr(\bm{W}(0),\bm{W}(1),\bm{Y}(0),\bm{Y}(1)|\bm{X})Pr(\bm{Z}|\bm{X})}
{\int Pr(\bm{W}(0),\bm{W}(1),\bm{Y}(0),\bm{Y}(1)|\bm{X})Pr(\bm{Z}|\bm{X})d\bm{W}^{mis}d\bm{Y}^{mis}}\\
=&\frac{Pr(\bm{W}(0),\bm{W}(1),\bm{Y}(0),\bm{Y}(1)|\bm{X})}
{\int Pr(\bm{W}(0),\bm{W}(1),\bm{Y}(0),\bm{Y}(1)|\bm{X})d\bm{W}^{mis}d\bm{Y}^{mis}}.
\end{aligned}\label{Bayesian_eq2}
\end{equation}
This implies that we can use a model to characterize the distribution \\$Pr(\bm{W}(0),\bm{W}(1),\bm{Y}(0),\bm{Y}(1)|\bm{X})$, and impute $\bm{W}^{mis}$ and $\bm{Y}^{mis}$ from the model.  The model is not regarded as being true, but is rather used as a tool to impute the missing values, and then to estimate sample CACE based on the imputed values.  If the Bayesian model is a good description of the process that has generated $\bm{W}(0)$, $\bm{W}(1)$, $\bm{Y}(0)$ and $\bm{Y}(1)$, when $\bm{W}(0)$, $\bm{W}(1)$, $\bm{Y}(0)$, $\bm{Y}(1)$ and $\bm{X}$ are fixed and
$\bm{Z}$ varies to reveal part of the potential indicators of treatment received and potential outcomes, $\bm{W}^{obs}$ and $\bm{Y}^{obs}$, we expect that the missing values can be predicted well and thus Bayesian inference can have good performance.

Suppose that we obtain $H$ posterior draws of missing values, $W_i^{mis\ (h)}$ and $Y_i^{mis\ (h)}$, $h=1,\cdots,H$.  Let $I_{i,co}^{(h)}$ denote the $h$th posterior draw of the indicator for whether unit $i$ is a complier based on $W_i^{obs}$ and $W_i^{mis\ (h)}$.  The $h$th posterior draw of $\tau_{CACE}^{samp}$ can be calculated as
	\begin{equation}
\begin{aligned}
		\tau_{CACE}^{samp\ (h)}&=\frac{\sum_{i=1}^n I^{(h)}_{i, co}\left(I\left(W_i^{obs}=1\right)\left(Y^{obs}_i-Y^{mis\ (h)}_i\right)+I\left(W_i^{obs}=0\right)\left(Y^{mis\ (h)}_i-Y^{obs}_i\right)\right)}{\sum_{i=1}^n I^{(h)}_{i, co}}\\
&=\frac{\sum_{i=1}^n I^{(h)}_{i, co}\left(2W_i^{obs}-1\right)\left(Y^{obs}_i-Y^{mis\ (h)}_i\right)}{\sum_{i=1}^n I^{(h)}_{i, co}}.
\end{aligned}\label{tau_postsamp}
	\end{equation}

\subsection{A Specific Bayesian Model}
\cite{imbens1997bayesian} proposed a Bayesian model to make inference about population CACE in randomized experiments with noncompliance.
They first model the distribution of $G_i$ conditional on $\bm{x}_i$, and then model the distribution of $(Y_i(1), Y_i(0))$ conditional on $G_i$ and $\bm{x}_i$. With the monotonicity assumption, for individuals with $Z_i = 1$ and $W_i=1$, $G_i$ could be $at$ or $co$, and for individuals with $Z_i=0$ and $W_i=0$, $G_i$ could be $nt$ or $co$.  Hence the observed-data likelihood has a complicated mixture structure. The posterior distribution can be sensitive to the choice of prior distribution \citep{hirano2000assessing}. This sensitivity is especially serious if $p_{co}$ is small.
	
We propose a different Bayesian approach, which invokes a commonly used	trick in modeling discrete variables \citep{albert1993bayesian}. We introduce latent variables $(L_i(0), L_i(1))$, where $W_{i}(z)=I(L_{i}(z)>0)$ for $z=0, 1$. Let $\bm{L}(z)$, $z=0,1$, denote the vector of $L_i(z)$ for all units. Let $\bm{\Theta}$ denote the set of model parameters. The joint distribution of $(\bm{Y}(0), \bm{Y}(1), \bm{W}(0), \bm{W}(1))$ given $\bm{X}$ and parameters $\bm{\Theta}$ can be written as
	\begin{equation}
		\begin{aligned}
			& Pr(\bm{Y}(0), \bm{Y}(1), \bm{W}(0), \bm{W}(1) \mid \bm{X}, \bm{\Theta}) \\
			=& \int Pr(\bm{Y}(0), \bm{Y}(1), \bm{W}(0), \bm{W}(1), \bm{L}(0), \bm{L}(1) \mid \bm{X}, \bm{\Theta}) d \bm{L}(0) d \bm{L}(1).
		\end{aligned}
	\end{equation}

	Since $(\bm{W}(0), \bm{W}(1))$ is determined by $(\bm{L}(0), \bm{L}(1))$, we can just model the distribution of $(\bm{Y}(0), \bm{Y}(1), \bm{L}(0), \bm{L}(1))$ given $\bm{X}$ and $\bm{\Theta}$. Let
	\begin{equation}
		\begin{aligned}
			& Pr(\bm{Y}(0), \bm{Y}(1), \bm{L}(0), \bm{L}(1) \mid \bm{X}, \bm{\Theta}) \\
			=& \prod_{i=1}^n f\left(Y_i(0), Y_i(1), L_i(0), L_i(1) \mid \bm{x}_i, \bm{\Theta}\right).
		\end{aligned}
	\end{equation}
	With the monotonicity assumption, $W_i(1) \geq W_i(0)$, we can make further simplification by letting $L_i(1) = L_i(0) + \alpha$ with $\alpha > 0$.
	
	The joint model of $(Y_i(0), Y_i(1), L_i(0))$ is taken to be a multivariate normal distribution with mean linearly depending on $\bm{x}_i$, i.e.
	\begin{equation}
		\begin{aligned}
			&Y_i(0)=\gamma_{00}+\bm{\gamma}_0^{\top} \bm{x}_i+\varepsilon_{i 0}, \\
			&Y_i(1)=\gamma_{10}+\bm{\gamma}_1^{\top} \bm{x}_i+\varepsilon_{i 1}, \\
			&L_i(0)=\beta_0+\bm{\beta}^{\top} \bm{x}_i+e_i,
		\end{aligned}
	\end{equation}
	where
	\begin{equation}
		\label{err}
		\left(\begin{array}{l}
			\varepsilon_{i 0} \\
			\varepsilon_{i 1} \\
			e_i
		\end{array}\right) \sim N\left(0,\left(\begin{array}{ccc}
			\sigma_0^2 & * & \pi_0 \\
			* & \sigma_1^2 & \pi_1 \\
			\pi_0 & \pi_1 & 1
		\end{array}\right)\right).
	\end{equation}
	For identifiability, the variance of $L_i(0)$ is set to be 1. The covariance of $Y_i(0)$ and $Y_i(1)$ is not identifiable since $(Y_i(0), Y_i(1))$ can never be both observed. The prior distribution, the algorithm for fitting the Bayesian model and imputing the missing values are given in the Appendix.

A related Bayesian model in the literature (e.g. \cite{lopes2014bayesian}) uses a common error term for $Y_i(0)$ and $Y_i(1)$ (i.e., setting $\varepsilon_{i 0}=\varepsilon_{i 1}$). With this, a constant treatment effect $Y_i(1)-Y_i(0)$ for individuals with the same value of $\bm{x}_i$ is assumed, ruling out unobserved heterogeneity and therefore may not be appropriate in some applications. Our Bayesian model instead allows unobserved heterogeneity and is more robust to model misspecification.

\section{Monte Carlo Simulation}
	The focus of the Monte Carlo simulation is to compare the performance of the methods under both CRE and ReM.  We use ``Wald" to denote the Wald method.  We use ``Reg" to denote the regression adjustment method, where the standard errors can be estimated using EHW, HC2 or HC3 estimators.  We use ``Bayes" to denote the Bayesian approach.  We run 4 chains with 2,500 posterior draws in each chain, discard the first 1,250 draws from each chain, and mix the remaining 5,000 draws for inference.  Posterior means are used as point estimates, and $95\%$ credible intervals are constructed using the $.025$ and $.975$ quantiles of posterior draws.

\subsection{Setup of the Monte Carlo Simulation}
In CRE, $n/2$ are randomly assigned to treatment, and the remaining $n/2$ are assigned to control.  In ReM, treatment assignment is randomized until the Mahalanobis distance is less than $\chi_{K, 0.01}^{2}$, the 0.01 quantile of a $\chi^{2}$ distribution with $K$ degrees of freedom.
	
	Each covariate independently follow a $N(0, 1)$ distribution. The potential outcomes, the potential latent variables and the potential indicators of treatment received are generated using the following data generating process:
	$$
	\begin{aligned}
		Y_i(0) &=\bm{\xi}^{\top}\bm{x}_i+\bm{\phi}^{\top}\bm{x}_i^2+\epsilon_{0 i}, \\
		Y_i(1) &=\bm{\xi}^{\top}\bm{x}_i+\bm{\eta}^{\top}\bm{x}_i+\bm{\phi}^{\top}\bm{x}_i^2+\epsilon_{1 i}, \\
		L_i(0) &=\delta_0+\bm{\psi}^{\top}\bm{x}_i+\bm{\phi}^{\top}\bm{x}_i^2+u_i, \\
		L_i(1) &=\delta_1+L_i(0), \\
		W_i(0) &=I\left(L_i(0)>0\right), \\
		W_i(1) &=I\left(L_i(1)>0\right)
	\end{aligned}
	$$
	Here $\bm{\xi}$, $\bm{\eta}$ and $\bm{\psi}$ are each a $K\times 1$ vector with all elements being 1,
$\bm{x}_i^2$ is $K\times 1$ vector containing elementwise square terms of $\bm{x}_i$.  Under a linear model (DGP1),
$\bm{\phi}$ is a $K\times 1$ vector with all elements being $\phi=0$.  Under a nonlinear model (DGP2), $\bm{\phi}$ is a $K\times 1$ vector with all elements being $\phi=0.4$.  For a given value of $p_{co}$, we set $\delta_0=((p_{co}-0.5)/0.35+1)(1-\phi)\sqrt{K}$, and choose the value of $\delta_1$ such that the fraction of compliers equals $p_{co}$.  We consider $n\in \{200,400\}$, $K\in \{5,10\}$, and three levels of fraction of compliers: $p_{co}\in \{0.85,0.5,0.15\}$.

We also consider four cases for the error terms. In the first two cases, the error terms are generated according to
	$$
	\left(\begin{array}{c}
		\kappa_0\epsilon_{i 0} \\
		\kappa_1\epsilon_{i 1} \\
		\kappa_2u_{i}
	\end{array}\right) \sim N\left(\bm{0},\left(\begin{array}{ccc}
		1 & 0 & \rho \\
		0 & 1 & \rho \\
		\rho & \rho & 1
	\end{array}\right)\right)
	$$
with $\rho=0$ in case 1 and $\rho=0.5$ in case 2.  The values of $\kappa_0$, $\kappa_1$ and $\kappa_2$ are chosen such that the squared multiple correlation coefficient is $0.5$ in each of the equations for $Y_i(0)$, $Y_i(1)$ and $L_i(0)$.  In the last two case, the error terms follow mixtures of normal distributions and centered exponential distributions, where the centered exponential distributions account for 20\% of the variances of the error terms, and the covariance matrices are respectively the same as in the first two cases.  The Bayesian model is correctly specified under DGP1 with the first two cases of error terms, and is incorrectly specified in the other scenarios.

The values of $(\bm{x}_i, L_i(0), L_i(1), W_i(0), W_i(1), Y_i(0), Y_i(1))$ are fixed, and only the values of $Z_i$ are randomly generated $1,000$ times under CRE or ReM. The observed values are $(\bm{x}_i, Z_i, W_i^{obs}, Y_i^{obs})$.  We thus have $1,000$ datasets for each DGP and each of 96 setting with different combinations of $n$, $K$, $p_{co}$, error distribution and randomization scheme.

	\subsection{Performance Measures}

For the Wald method or the regression adjustment method, $\widehat{ITT}_W$ or $\widehat{ITT}_W^{adj}$ can be negative, indicating an unreasonable negative estimate of the fraction of compliers.  For example, under DGP1 with $p_{co}=0.15$, $n=200$, $K=10$, case 1 error distribution and CRE, for one of the 1,000 datasets, $\widehat{ITT}_W$ is negative; under DGP1 with $p_{co}=0.15$, $n=200$, $K=5$, case 1 error distribution and ReM, for one of the 1,000 datasets, $\widehat{ITT}_W^{adj}$ is negative.  We remove such results in our comparison.

\begin{figure}
  \centering
  \subfigure[point estimates]{\includegraphics[width=10cm]{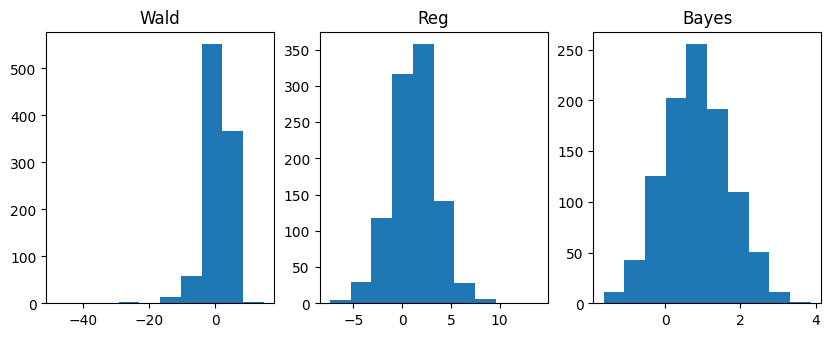}}
  \subfigure[lengths of 95\% intervals]{\includegraphics[width=10cm]{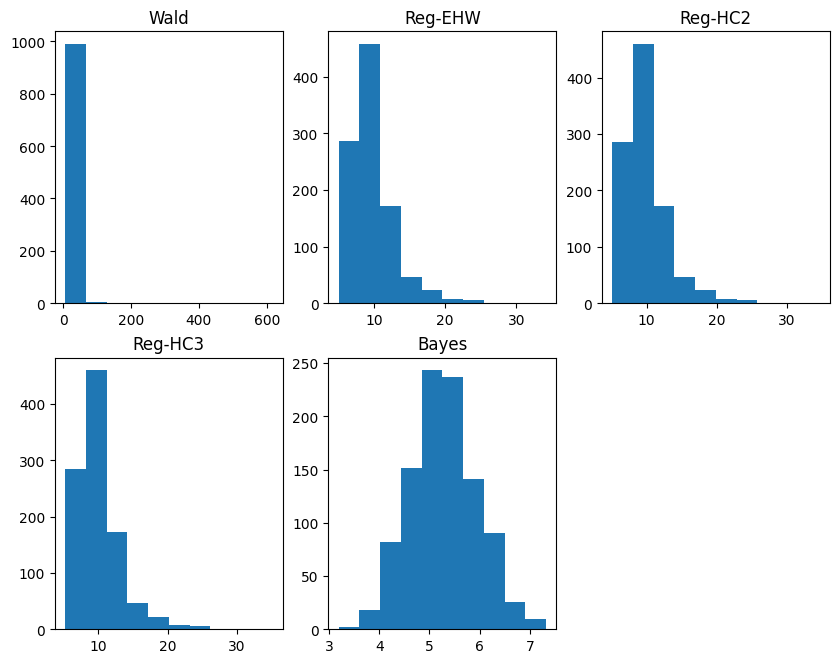}}
  \caption{Histograms of point estimates and lengths of 95\% intervals under DGP1 with $p_{co}=0.15$, $n=200$, $K=10$ and case 1 error distributions.}
  \label{fig:hist}
\end{figure}

 Another problem with the Wald method or the regression adjustment method is instability.  When $\widehat{ITT}_W$ or $\widehat{ITT}_W^{adj}$ is close to zero, the point estimate given by the Wald method or the regression adjustment method can be quite large in absolute value, and the corresponding interval estimates can be quite wide.  To illustrate this point, for the data sets generated under DGP1 with $p_{co}=0.15$, $n=200$, $K=10$, case 1 error distributions and CRE, Figure~\ref{fig:hist} presents histograms of point estimates and lengths of 95\% intervals.  It shows the Wald method and the regression adjustment method can be unstable, whereas the Bayesian method is stable.

For further comparison, we compare the performance of the methods using median absolute error and median length of 95\% intervals.  We also compare the coverage rate of 95\% intervals. Specifically, for the $r$th dataset and the $m$th method, let $\widehat{\tau}_{r m}$ denote the point estimate, let $\mathcal{C}_{r m}$ denote the $95\%$ interval estimate, and let $L_{r m}$ denote the length of $\mathcal{C}_{r m}$. The performance measures for method $m$ are
\begin{equation}
	\begin{aligned}
		\text{MAE}_{m} &=\text{median of }\{|\widehat{\tau}_{r m}-\tau_{CACE}^{samp}|,\ r=1,\cdots,1000\}, \\
		\text{CRate}_{m} &=\frac{1}{1,000} \sum_{r=1}^{1,000} I\left(\tau_{CACE}^{samp} \in \mathcal{C}_{r m}\right),\\
\text{Len}_m &= \text{median of }\{L_{rm},\ r=1,\cdots,1000\}.
	\end{aligned}
\end{equation}
	
\subsection{Results}

For the performance of different methods under DGP1, Table~\ref{T:MAE:DGP1} presents median absolute error, Table~\ref{T:CR:DGP1} presents coverage rate of 95\% intervals measured in percentage differences from the nominal level 0.95, $100\times (\text{CRate}_m-0.95)$.  Table~\ref{T:LEN:DGP1} presents median length of 95\% intervals.  The corresponding results for DGP2 are presented in Tables~\ref{T:MAE:DGP2}-~\ref{T:LEN:DGP2}.

\begin{landscape}
	\begin{table}[tbp]
		\begin{scriptsize}
			\caption{Median absolute error for different methods under DGP1.}
			\label{T:MAE:DGP1}
			\begin{center}
				\begin{tabular}{llS[table-format=1.3]S[table-format=1.3]S[table-format=1.3]S[table-format=1.3]S[table-format=1.3]S[table-format=1.3]S[table-format=1.3]S[table-format=1.3]S[table-format=1.3]S[table-format=1.3]S[table-format=1.3]S[table-format=1.3]}
					\hline\hline
					& & \multicolumn{4}{c}{$p_{co}=0.85$} & \multicolumn{4}{c}{$p_{co}=0.5$} & \multicolumn{4}{c}{$p_{co}=0.15$}\\
					method & rand & {case 1} & {case 2} & {case 3} & {case 4} & {case 1} & {case 2} & {case 3} & {case 4} & {case 1} & {case 2} & {case 3} & {case 4}\\
					\hline
					& &\multicolumn{12}{c}{$K=5,\ n=200$}\\
					\multirow{2}{*}{Wald} &CRE     &  0.488 &  0.540 &  0.505 &  0.551 &  0.947 &  0.988 &  0.937 &  0.936 &  3.332 &  3.330 &  2.992 &  2.949 \\
					&ReM     &  0.357 &  0.390 &  0.342 &  0.367 &  0.711 &  0.729 &  0.695 &  0.720 &  2.580 &  2.474 &  2.508 &  2.367 \\
					\multirow{2}{*}{Reg} &CRE&  0.349 &  0.387 &  0.359 &  0.363 &  0.716 &  0.742 &  0.704 &  0.682 &  2.431 &  2.550 &  2.258 &  2.299 \\
					&ReM &  0.340 &  0.373 &  0.325 &  0.365 &  0.690 &  0.695 &  0.676 &  0.693 &  2.458 &  2.316 &  2.393 &  2.214 \\
					\multirow{2}{*}{Bayes} &CRE &  0.340 &  0.368 &  0.353 &  0.334 &  0.653 &  0.641 &  0.646 &  0.528 &  1.476 &  0.767 &  1.058 &  0.576 \\
					& ReM   &  0.333 &  0.357 &  0.338 &  0.321 &  0.644 &  0.633 &  0.622 &  0.481 &  1.452 &  0.791 &  1.037 &  0.580 \\
					\hline
					& &\multicolumn{12}{c}{$K=5,\ n=400$}\\
					\multirow{2}{*}{Wald} &CRE     &  0.329 &  0.339 &  0.347 &  0.369 &  0.600 &  0.598 &  0.592 &  0.627 &  1.895 &  1.937 &  1.946 &  1.914 \\
					&ReM     &  0.229 &  0.262 &  0.228 &  0.238 &  0.433 &  0.454 &  0.450 &  0.452 &  1.468 &  1.537 &  1.476 &  1.542 \\
					\multirow{2}{*}{Reg} &CRE&  0.226 &  0.251 &  0.221 &  0.242 &  0.399 &  0.452 &  0.456 &  0.428 &  1.383 &  1.457 &  1.426 &  1.465 \\
					&ReM &  0.219 &  0.253 &  0.208 &  0.235 &  0.432 &  0.450 &  0.431 &  0.444 &  1.448 &  1.513 &  1.466 &  1.581 \\
					\multirow{2}{*}{Bayes} &CRE     &  0.207 &  0.243 &  0.201 &  0.223 &  0.396 &  0.444 &  0.381 &  0.370 &  0.621 &  0.731 &  0.656 &  1.348 \\
					& ReM     &  0.210 &  0.244 &  0.198 &  0.216 &  0.417 &  0.483 &  0.369 &  0.385 &  0.647 &  0.784 &  0.591 &  1.370 \\
					\hline
					& &\multicolumn{12}{c}{$K=10,\ n=200$}\\
					\multirow{2}{*}{Wald} &CRE      &  0.680 &  0.743 &  0.668 &  0.757 &  1.289 &  1.335 &  1.253 &  1.346 &  4.280 &  4.364 &  4.449 &  4.040 \\
					&ReM    &  0.510 &  0.592 &  0.541 &  0.619 &  0.979 &  1.056 &  1.043 &  1.049 &  3.620 &  3.671 &  3.821 &  3.296 \\
					\multirow{2}{*}{Reg} &CRE &  0.486 &  0.544 &  0.458 &  0.550 &  0.908 &  0.981 &  0.925 &  1.008 &  3.105 &  3.328 &  3.376 &  3.301 \\
					&ReM&  0.458 &  0.552 &  0.498 &  0.551 &  0.917 &  0.983 &  0.949 &  1.002 &  3.348 &  3.499 &  3.542 &  3.080 \\
					\multirow{2}{*}{Bayes} &CRE     &  0.436 &  0.480 &  0.431 &  0.558 &  0.703 &  0.793 &  0.696 &  0.857 &  1.511 &  2.788 &  1.327 &  1.705 \\
					&ReM    &  0.418 &  0.498 &  0.470 &  0.535 &  0.748 &  0.877 &  0.687 &  0.863 &  1.653 &  2.829 &  1.481 &  1.800 \\
					\hline
					& &\multicolumn{12}{c}{$K=10,\ n=400$}\\
					\multirow{2}{*}{Wald} &CRE     &  0.503 &  0.529 &  0.502 &  0.512 &  0.897 &  0.952 &  0.896 &  0.950 &  2.829 &  2.977 &  2.882 &  2.956 \\
					&ReM     &  0.376 &  0.433 &  0.362 &  0.419 &  0.696 &  0.722 &  0.673 &  0.818 &  2.287 &  2.384 &  2.222 &  2.678 \\
					\multirow{2}{*}{Reg} &CRE &  0.333 &  0.368 &  0.333 &  0.328 &  0.654 &  0.663 &  0.640 &  0.605 &  2.126 &  2.124 &  2.166 &  2.200 \\
					&ReM &  0.321 &  0.368 &  0.303 &  0.369 &  0.618 &  0.659 &  0.607 &  0.727 &  2.095 &  2.139 &  2.005 &  2.438 \\
					\multirow{2}{*}{Bayes} &CRE    &  0.308 &  0.347 &  0.310 &  0.325 &  0.568 &  0.600 &  0.578 &  0.574 &  1.808 &  1.588 &  1.198 &  0.637 \\
					&ReM    &  0.295 &  0.336 &  0.289 &  0.365 &  0.563 &  0.568 &  0.526 &  0.593 &  1.750 &  1.560 &  1.157 &  0.702 \\
					\hline
					\hline				
				\end{tabular}
			\end{center}
		\end{scriptsize}
	\end{table}
\end{landscape}
	
\begin{landscape}
	\begin{table}[tbp]
		\begin{scriptsize}
			\caption{Coverage rate of 95\% intervals for different methods under DGP1, measured in percentage differences from the nominal level 0.95, $100\times (\text{CRate}_m-0.95)$.}
			\label{T:CR:DGP1}
			\begin{center}
				\begin{tabular}{llS[table-format=1.3]S[table-format=1.3]S[table-format=1.3]S[table-format=1.3]S[table-format=1.3]S[table-format=1.3]S[table-format=1.3]S[table-format=1.3]S[table-format=1.3]S[table-format=1.3]S[table-format=1.3]S[table-format=1.3]}
					\hline\hline
					& &\multicolumn{4}{c}{$p_{co}=0.85$} & \multicolumn{4}{c}{$p_{co}=0.5$} & \multicolumn{4}{c}{$p_{co}=0.15$}\\
					method & rand &{case 1} & {case 2} & {case 3} & {case 4} & {case 1} & {case 2} & {case 3} & {case 4} & {case 1} & {case 2} & {case 3} & {case 4}\\
					\hline
					& &\multicolumn{12}{c}{$K=5,\ n=200$}\\
					\multirow{2}{*}{Wald} &CRE      &  2.5 &  1.8 &   2.2 &  1.1 &  2.0 &  1.4 &  1.5 &  1.2 &  2.1 &  1.6 &  3.1 &  3.1 \\
					&ReM      &  3.1 &  1.7 &   2.7 &  0.9 &  0.9 &  0.5 &  0.4 &  0.7 &  4.4 &  3.8 &  4.0 &  4.1 \\
					\multirow{2}{*}{Reg-EHW} &CRE &  2.9 &  1.2 &   2.5 &  1.5 &  1.8 &  0.3 &  2.4 &  1.0 &  4.0 &  4.1 &  4.0 &  3.7 \\
					&ReM &  3.6 &  2.3 &   3.1 &  1.6 &  1.6 &  1.3 &  1.0 &  1.4 &  4.3 &  4.2 &  4.0 &  4.0 \\
					\multirow{2}{*}{Reg-HC2} &CRE  &  3.3 &  1.8 &   2.8 &  2.2 &  2.4 &  1.5 &  2.7 &  1.9 &  4.3 &  4.3 &  4.1 &  4.0 \\
					&ReM &  3.9 &  2.7 &   3.4 &  2.4 &  2.1 &  2.1 &  1.6 &  2.1 &  4.5 &  4.6 &  4.4 &  4.2 \\
					\multirow{2}{*}{Reg-HC3} &CRE &  3.7 &  2.6 &   3.3 &  2.7 &  3.2 &  1.8 &  2.9 &  2.1 &  4.4 &  4.6 &  4.4 &  4.3 \\
					&ReM &  4.3 &  3.1 &   3.5 &  2.9 &  2.8 &  2.6 &  2.1 &  2.4 &  4.8 &  4.7 &  4.4 &  4.3 \\
					\multirow{2}{*}{Bayes} &CRE      &  2.0 &  1.5 &  -1.4 &  2.6 &  2.4 &  1.1 &  1.8 &  2.2 &  3.4 &  4.7 &  4.2 &  4.7 \\
					&ReM     &  3.0 &  1.3 &  -0.4 &  1.1 &  1.4 &  1.5 &  0.6 &  2.2 &  2.9 &  4.8 &  3.9 &  4.9 \\
					\hline
					& &\multicolumn{12}{c}{$K=5,\ n=400$}\\
					\multirow{2}{*}{Wald} &CRE      &  2.7 &  1.7 &  2.6 &  1.6 &  1.6 &  1.4 &  1.8 &  1.0 &  2.3 &  2.3 &  2.7 &   2.3 \\
					&ReM     &  3.8 &  2.8 &  3.4 &  2.7 &  1.8 &  0.9 &  2.3 &  1.2 &  3.0 &  2.5 &  2.7 &   2.7 \\
					\multirow{2}{*}{Reg-EHW} &CRE  &  3.7 &  2.2 &  3.5 &  3.1 &  3.1 &  1.6 &  2.8 &  2.1 &  3.8 &  3.2 &  3.6 &   2.7 \\
					&ReM &  4.3 &  2.7 &  4.1 &  3.4 &  3.4 &  2.1 &  2.2 &  2.5 &  3.3 &  3.1 &  2.8 &   3.1 \\
					\multirow{2}{*}{Reg-HC2} &CRE  &  4.1 &  2.4 &  3.8 &  3.5 &  3.2 &  1.7 &  3.0 &  2.3 &  3.8 &  3.5 &  3.9 &   2.9 \\
					&ReM&  4.3 &  2.8 &  4.3 &  3.4 &  3.5 &  2.3 &  2.4 &  2.8 &  3.6 &  3.2 &  3.1 &   3.2 \\
					\multirow{2}{*}{Reg-HC3} &CRE &  4.3 &  2.5 &  3.9 &  3.7 &  3.4 &  1.9 &  3.3 &  2.5 &  4.0 &  3.6 &  4.1 &   3.2 \\
					&ReM&  4.3 &  3.0 &  4.3 &  3.5 &  3.6 &  2.8 &  2.6 &  3.0 &  3.7 &  3.3 &  3.5 &   3.3 \\
					\multirow{2}{*}{Bayes} &CRE      &  3.8 &  1.3 &  3.2 &  2.6 &  2.9 &  1.6 &  3.2 &  1.9 &  3.8 &  3.2 &  4.6 &  -2.3 \\
					&ReM    &  4.0 &  2.6 &  4.0 &  1.8 &  3.3 &  1.8 &  2.9 &  1.2 &  4.1 &  3.9 &  4.1 &  -2.5 \\
					\hline
					& &\multicolumn{12}{c}{$K=10,\ n=200$}\\
					\multirow{2}{*}{Wald} &CRE       &  1.0 &   1.2 &  2.2 &  1.1 &   1.3 &   0.7 &  1.3 &   2.0 &  1.7 &   2.1 &  1.4 &  3.3 \\
					&ReM     &  3.3 &   1.7 &  2.9 &  3.2 &   2.7 &   1.9 &  2.7 &   2.4 &  3.4 &   4.1 &  3.8 &  4.4 \\
					\multirow{2}{*}{Reg-EHW} &CRE &  2.9 &   0.8 &  2.5 &  1.3 &  1.3 &  -0.6 &  1.7 &  0.9 &  4.2 &   3.8 &  4.1 &  3.8 \\
					&ReM &  2.4 &  -0.3 &  2.3 &  2.0 &  0.0 &   0.0 &  1.2 &  1.5 &  3.6 &   3.6 &  3.0 &  3.9 \\
					\multirow{2}{*}{Reg-HC2} &CRE  &  3.4 &   1.9 &  3.4 &  2.3 &  2.0 &   1.0 &  2.8 &  2.0 &  4.8 &   4.2 &  4.6 &  4.3 \\
					&ReM &  3.0 &   0.9 &  3.0 &  3.4 &  1.2 &   1.1 &  2.5 &  2.4 &  4.3 &   4.6 &  3.7 &  4.4 \\
					\multirow{2}{*}{Reg-HC3} &CRE  &  3.8 &   2.9 &  4.4 &  3.3 &  2.8 &   2.5 &  3.7 &  2.8 &  4.9 &   4.8 &  4.8 &  4.5 \\
					&ReM &  3.5 &   2.6 &  3.5 &  4.2 &  2.6 &   2.4 &  2.9 &  3.4 &  4.5 &   4.7 &  4.5 &  4.7 \\
					\multirow{2}{*}{Bayes} &CRE     &  2.5 &   0.8 &  3.3 &  0.9 &   3.3 &   2.3 &  3.8 &   2.7 &  2.6 &  -2.6 &  3.1 &  3.1 \\
					&ReM    &  2.1 &   1.1 &  2.6 &  1.9 &   2.7 &   1.6 &  3.2 &   2.3 &  2.2 &  -2.9 &  2.3 &  2.8 \\
					\hline
					& & \multicolumn{12}{c}{$K=10,\ n=400$}\\
					
					\multirow{2}{*}{Wald} &CRE       &  2.4 &  1.2 &  2.1 &  2.1 &  1.8 &  1.5 &  1.6 &   1.5 &   1.9 &  1.6 &  2.9 &  1.2 \\
					&ReM     &  3.8 &  2.4 &  3.8 &  1.9 &  3.0 &  2.5 &  2.5 &   1.1 &   3.1 &  3.0 &  3.5 &  2.7 \\
					\multirow{2}{*}{Reg-EHW} &CRE  &  2.8 &  1.4 &  3.1 &  2.9 &  1.6 &  0.7 &  2.2 &  1.9 &   3.1 &  2.8 &  2.7 &  2.7 \\
					&ReM &  3.6 &  1.4 &  3.5 &  1.4 &  1.6 &  0.9 &  1.5 &  0.0 &   2.4 &  3.4 &  3.0 &  2.8 \\
					\multirow{2}{*}{Reg-HC2} &CRE  &  3.1 &  1.9 &  3.2 &  3.3 &  2.1 &  1.3 &  2.5 &  2.4 &   3.3 &  3.1 &  3.3 &  3.2 \\
					&ReM &  3.7 &  1.7 &  3.8 &  2.3 &  2.2 &  1.5 &  1.9 &  0.9 &   3.1 &  3.7 &  3.2 &  3.1 \\
					\multirow{2}{*}{Reg-HC3} &CRE &  3.5 &  2.4 &  3.6 &  3.6 &  2.7 &  1.9 &  2.6 &  3.1 &   3.8 &  3.5 &  3.9 &  3.6 \\
					&ReM &  4.0 &  2.4 &  3.9 &  2.8 &  2.3 &  1.8 &  2.2 &  1.9 &   3.3 &  3.8 &  3.6 &  3.7 \\
					\multirow{2}{*}{Bayes} &CRE     &  2.4 &  1.4 &  3.1 &  2.1 &  1.4 &  1.0 &  2.1 &   2.1 &   0.6 &  0.3 &  3.5 &  4.8 \\
					&ReM    &  3.5 &  1.2 &  3.2 &  0.6 &  1.9 &  1.1 &  2.9 &   1.4 &  -0.1 &  0.6 &  3.4 &  5.0 \\
					\hline\hline
				\end{tabular}
			\end{center}
		\end{scriptsize}
	\end{table}
\end{landscape}	

\begin{landscape}
	\begin{table}[tbp]
		\begin{scriptsize}
			\caption{Median length of 95\% interval for different methods under DGP1.}
			\label{T:LEN:DGP1}
			\begin{center}
				\begin{tabular}{llS[table-format=1.3]S[table-format=1.3]S[table-format=1.3]S[table-format=1.3]S[table-format=1.3]S[table-format=1.3]S[table-format=1.3]S[table-format=1.3]S[table-format=1.3]S[table-format=1.3]S[table-format=1.3]S[table-format=1.3]}
					\hline\hline
					& & \multicolumn{4}{c}{$p_{co}=0.85$} & \multicolumn{4}{c}{$p_{co}=0.5$} & \multicolumn{4}{c}{$p_{co}=0.15$}\\
					method & rand & {case 1} & {case 2} & {case 3} & {case 4} & {case 1} & {case 2} & {case 3} & {case 4} & {case 1} & {case 2} & {case 3} & {case 4}\\
					\hline
					& &\multicolumn{12}{c}{$K=5,\ n=200$}\\
					\multirow{2}{*}{Wald} &CRE       &  3.332 &  3.346 &  3.376 &  3.318 &  5.971 &  6.000 &  5.923 &  5.767 &  20.292 &  20.105 &  18.714 &  18.049 \\
					&ReM       &  2.436 &  2.422 &  2.454 &  2.255 &  4.337 &  4.244 &  4.300 &  4.082 &  14.778 &  14.141 &  14.441 &  13.631 \\
					\multirow{2}{*}{Reg-EHW} &CRE   &  2.458 &  2.440 &  2.469 &  2.273 &  4.376 &  4.284 &  4.339 &  4.128 &  15.196 &  14.457 &  14.524 &  13.354 \\
					&ReM  &  2.431 &  2.415 &  2.448 &  2.247 &  4.319 &  4.223 &  4.299 &  4.070 &  14.767 &  14.149 &  14.262 &  13.570 \\
					\multirow{2}{*}{Reg-HC2} &CRE   &  2.540 &  2.523 &  2.551 &  2.346 &  4.526 &  4.431 &  4.493 &  4.257 &  15.729 &  14.965 &  14.999 &  13.788 \\
					&ReM &  2.512 &  2.494 &  2.533 &  2.320 &  4.466 &  4.368 &  4.453 &  4.198 &  15.263 &  14.629 &  14.737 &  13.988 \\
					\multirow{2}{*}{Reg-HC3} &CRE   &  2.627 &  2.609 &  2.641 &  2.424 &  4.684 &  4.583 &  4.655 &  4.393 &  16.280 &  15.500 &  15.529 &  14.262 \\
					&ReM  &  2.598 &  2.579 &  2.621 &  2.397 &  4.624 &  4.518 &  4.613 &  4.332 &  15.786 &  15.136 &  15.252 &  14.423 \\
					\multirow{2}{*}{Bayes} &CRE     &  2.309 &  2.292 &  2.129 &  2.087 &  4.171 &  4.009 &  3.989 &  3.285 &   7.946 &   7.221 &   8.288 &   6.944 \\
					&ReM     &  2.285 &  2.261 &  2.088 &  2.048 &  4.102 &  3.951 &  3.906 &  3.245 &   7.915 &   7.177 &   8.182 &   6.975 \\
					\hline
					& &\multicolumn{12}{c}{$K=5,\ n=400$}\\
					\multirow{2}{*}{Wald} &CRE       &  2.260 &  2.215 &  2.293 &  2.325 &  3.832 &  3.784 &  3.886 &  3.881 &  11.819 &  11.996 &  12.215 &  12.181 \\
					&ReM     &  1.684 &  1.662 &  1.681 &  1.611 &  2.855 &  2.845 &  2.890 &  2.748 &   8.907 &   8.957 &   8.936 &   8.885 \\
					\multirow{2}{*}{Reg-EHW} &CRE  &  1.677 &  1.656 &  1.676 &  1.606 &  2.857 &  2.851 &  2.898 &  2.737 &   8.985 &   9.138 &   9.148 &   8.871 \\
					&ReM  &  1.675 &  1.650 &  1.670 &  1.596 &  2.835 &  2.822 &  2.876 &  2.716 &   8.834 &   8.900 &   8.868 &   8.835 \\
					\multirow{2}{*}{Reg-HC2} &CRE   &  1.703 &  1.681 &  1.703 &  1.631 &  2.901 &  2.897 &  2.943 &  2.779 &   9.128 &   9.292 &   9.291 &   9.000 \\
					&ReM &  1.700 &  1.675 &  1.695 &  1.621 &  2.880 &  2.868 &  2.921 &  2.758 &   8.969 &   9.046 &   9.004 &   8.964 \\
					\multirow{2}{*}{Reg-HC3} &CRE   &  1.729 &  1.708 &  1.729 &  1.656 &  2.946 &  2.943 &  2.989 &  2.822 &   9.272 &   9.449 &   9.436 &   9.133 \\
					&ReM &  1.726 &  1.700 &  1.721 &  1.645 &  2.926 &  2.914 &  2.967 &  2.800 &   9.109 &   9.197 &   9.144 &   9.092 \\
					\multirow{2}{*}{Bayes} &CRE      &  1.526 &  1.514 &  1.502 &  1.438 &  2.642 &  2.714 &  2.552 &  2.380 &   5.268 &   5.827 &   5.403 &   5.610 \\
					&ReM      &  1.522 &  1.509 &  1.493 &  1.428 &  2.599 &  2.680 &  2.521 &  2.362 &   5.184 &   5.750 &   5.328 &   5.631 \\
					\hline
					& &\multicolumn{12}{c}{$K=10,\ n=200$}\\
					\multirow{2}{*}{Wald} &CRE       &  4.467 &  4.462 &  4.769 &  4.853 &  7.772 &  7.781 &  8.324 &  8.120 &  26.210 &  25.718 &  27.209 &  24.198 \\
					&ReM       &  3.623 &  3.678 &  3.865 &  4.035 &  6.367 &  6.488 &  6.801 &  6.816 &  22.105 &  22.225 &  23.438 &  21.027 \\
					\multirow{2}{*}{Reg-EHW} &CRE   &  3.089 &  3.208 &  3.330 &  3.598 &  5.389 &  5.594 &  5.740 &  6.106 &  18.497 &  18.654 &  19.484 &  18.233 \\
					&ReM &  3.041 &  3.159 &  3.254 &  3.506 &  5.333 &  5.545 &  5.663 &  5.970 &  18.564 &  18.539 &  19.382 &  18.148 \\
					\multirow{2}{*}{Reg-HC2} &CRE   &  3.287 &  3.413 &  3.535 &  3.826 &  5.719 &  5.952 &  6.097 &  6.474 &  19.640 &  19.823 &  20.655 &  19.382 \\
					&ReM  &  3.230 &  3.355 &  3.452 &  3.723 &  5.655 &  5.898 &  6.013 &  6.329 &  19.707 &  19.733 &  20.594 &  19.227 \\
					\multirow{2}{*}{Reg-HC3} &CRE   &  3.496 &  3.632 &  3.755 &  4.069 &  6.074 &  6.337 &  6.490 &  6.876 &  20.846 &  21.109 &  21.959 &  20.622 \\
					&ReM  &  3.435 &  3.568 &  3.669 &  3.959 &  5.998 &  6.271 &  6.397 &  6.713 &  20.912 &  21.009 &  21.873 &  20.446 \\
					\multirow{2}{*}{Bayes} &CRE       &  2.879 &  3.018 &  3.216 &  3.376 &  4.912 &  5.229 &  5.131 &  5.635 &   9.604 &  10.489 &  10.678 &  11.553 \\
					&ReM      &  2.834 &  2.949 &  3.149 &  3.306 &  4.869 &  5.155 &  5.065 &  5.516 &   9.587 &  10.587 &  10.615 &  11.428 \\
					\hline
					& &\multicolumn{12}{c}{$K=10,\ n=400$}\\
					\multirow{2}{*}{Wald} &CRE        &  3.215 &  3.344 &  3.209 &  3.266 &  5.414 &  5.543 &  5.378 &  5.714 &  17.172 &  17.453 &  16.569 &  18.026 \\
					&ReM      &  2.683 &  2.769 &  2.668 &  2.724 &  4.584 &  4.609 &  4.541 &  4.848 &  14.456 &  14.380 &  14.253 &  15.810 \\
					\multirow{2}{*}{Reg-EHW} &CRE   &  2.308 &  2.363 &  2.294 &  2.367 &  4.009 &  3.957 &  3.940 &  4.218 &  12.707 &  12.396 &  12.443 &  13.676 \\
					&ReM  &  2.280 &  2.334 &  2.269 &  2.345 &  3.974 &  3.912 &  3.905 &  4.180 &  12.613 &  12.258 &  12.461 &  13.851 \\
					\multirow{2}{*}{Reg-HC2} &CRE   &  2.372 &  2.430 &  2.357 &  2.438 &  4.124 &  4.069 &  4.052 &  4.343 &  13.076 &  12.752 &  12.800 &  14.087 \\
					&ReM  &  2.344 &  2.401 &  2.333 &  2.415 &  4.085 &  4.024 &  4.014 &  4.304 &  12.974 &  12.617 &  12.811 &  14.269 \\
					\multirow{2}{*}{Reg-HC3} &CRE  &  2.440 &  2.500 &  2.424 &  2.511 &  4.243 &  4.185 &  4.166 &  4.473 &  13.458 &  13.128 &  13.173 &  14.524 \\
					&ReM &  2.411 &  2.469 &  2.398 &  2.487 &  4.203 &  4.141 &  4.128 &  4.435 &  13.353 &  12.987 &  13.179 &  14.696 \\
					\multirow{2}{*}{Bayes} &CRE       &  2.145 &  2.153 &  2.136 &  2.232 &  3.717 &  3.625 &  3.631 &  3.744 &   8.914 &   8.513 &   8.193 &   6.848 \\
					&ReM      &  2.124 &  2.132 &  2.116 &  2.205 &  3.697 &  3.587 &  3.593 &  3.712 &   8.849 &   8.419 &   8.204 &   6.830 \\
					\hline\hline
				\end{tabular}
			\end{center}
		\end{scriptsize}
	\end{table}
\end{landscape}
	
\begin{landscape}
	\begin{table}[tbp]
		\begin{scriptsize}
			\caption{Median absolute error for different methods under DGP2.}
			\label{T:MAE:DGP2}
			\begin{center}
				\begin{tabular}{llS[table-format=1.3]S[table-format=1.3]S[table-format=1.3]S[table-format=1.3]S[table-format=1.3]S[table-format=1.3]S[table-format=1.3]S[table-format=1.3]S[table-format=1.3]S[table-format=1.3]S[table-format=1.3]S[table-format=1.3]}
					\hline\hline
					& & \multicolumn{4}{c}{$p_{co}=0.85$} & \multicolumn{4}{c}{$p_{co}=0.5$} & \multicolumn{4}{c}{$p_{co}=0.15$}\\
					method & rand & {case 1} & {case 2} & {case 3} & {case 4} & {case 1} & {case 2} & {case 3} & {case 4} & {case 1} & {case 2} & {case 3} & {case 4}\\
					\hline
					& &\multicolumn{12}{c}{$K=5,\ n=200$}\\
					\multirow{2}{*}{Wald} &CRE      &  0.539 &  0.573 &  0.547 &  0.606 &  1.089 &  1.164 &  1.093 &  1.092 &  3.851 &  3.958 &  3.779 &  3.635 \\
					&ReM     &  0.429 &  0.471 &  0.412 &  0.429 &  0.871 &  0.900 &  0.869 &  0.814 &  3.257 &  3.072 &  3.212 &  2.876 \\
					\multirow{2}{*}{Reg} &CRE  &  0.396 &  0.459 &  0.410 &  0.441 &  0.902 &  0.952 &  0.886 &  0.826 &  3.029 &  3.175 &  3.176 &  2.813 \\
					&ReM&  0.404 &  0.441 &  0.414 &  0.423 &  0.878 &  0.844 &  0.841 &  0.798 &  3.226 &  3.078 &  3.186 &  2.756 \\
					\multirow{2}{*}{Bayes} &CRE     &  0.403 &  0.463 &  0.389 &  0.408 &  0.775 &  0.802 &  0.722 &  0.646 &  1.596 &  1.146 &  1.427 &  0.770 \\
					&ReM    &  0.390 &  0.433 &  0.403 &  0.417 &  0.798 &  0.807 &  0.742 &  0.630 &  1.640 &  1.202 &  1.346 &  0.754 \\
					\hline
					& &\multicolumn{12}{c}{$K=5,\ n=400$}\\
					\multirow{2}{*}{Wald} &CRE      &  0.373 &  0.396 &  0.372 &  0.379 &  0.686 &  0.703 &  0.706 &  0.721 &  2.638 &  2.509 &  2.494 &  2.676 \\
					&ReM      &  0.258 &  0.298 &  0.273 &  0.291 &  0.554 &  0.554 &  0.563 &  0.523 &  2.074 &  2.094 &  2.047 &  1.947 \\
					\multirow{2}{*}{Reg} &CRE  &  0.271 &  0.292 &  0.263 &  0.276 &  0.532 &  0.541 &  0.512 &  0.511 &  1.933 &  1.912 &  1.889 &  1.855 \\
					&ReM&  0.257 &  0.297 &  0.260 &  0.271 &  0.535 &  0.575 &  0.547 &  0.527 &  2.016 &  1.993 &  1.979 &  1.969 \\
					\multirow{2}{*}{Bayes} &CRE     &  0.260 &  0.288 &  0.254 &  0.256 &  0.463 &  0.506 &  0.429 &  0.421 &  1.194 &  0.732 &  1.368 &  0.781 \\
					&ReM     &  0.263 &  0.291 &  0.258 &  0.261 &  0.504 &  0.516 &  0.462 &  0.427 &  1.224 &  0.705 &  1.327 &  0.820 \\
					\hline
					& &\multicolumn{12}{c}{$K=10,\ n=200$}\\
					\multirow{2}{*}{Wald} &CRE       &  0.817 &  0.870 &  0.752 &  0.892 &  1.563 &  1.585 &  1.495 &  1.696 &  5.792 &  5.812 &  5.320 &  6.225 \\
					&ReM    &  0.595 &  0.712 &  0.613 &  0.719 &  1.208 &  1.335 &  1.269 &  1.374 &  4.424 &  4.851 &  4.779 &  5.006 \\
					\multirow{2}{*}{Reg} &CRE  &  0.613 &  0.672 &  0.526 &  0.711 &  1.081 &  1.220 &  1.146 &  1.288 &  3.970 &  4.573 &  4.094 &  4.580 \\
					&ReM &  0.595 &  0.654 &  0.550 &  0.652 &  1.182 &  1.259 &  1.226 &  1.278 &  4.289 &  4.726 &  4.422 &  4.812 \\
					\multirow{2}{*}{Bayes} &CRE     &  0.554 &  0.637 &  0.546 &  0.659 &  0.964 &  1.112 &  0.841 &  1.143 &  1.533 &  2.078 &  3.397 &  2.326 \\
					&ReM    &  0.535 &  0.633 &  0.555 &  0.638 &  1.007 &  1.196 &  0.867 &  1.077 &  1.624 &  2.217 &  3.507 &  2.336 \\
					\hline
					& &\multicolumn{12}{c}{$K=10,\ n=400$}\\
					\multirow{2}{*}{Wald} &CRE      &  0.549 &  0.594 &  0.555 &  0.565 &  0.993 &  1.075 &  1.056 &  1.073 &  3.660 &  3.741 &  3.660 &  3.677 \\
					&ReM    &  0.416 &  0.475 &  0.401 &  0.479 &  0.782 &  0.815 &  0.815 &  0.926 &  2.932 &  2.916 &  2.871 &  3.236 \\
					\multirow{2}{*}{Reg} &CRE &  0.379 &  0.410 &  0.358 &  0.395 &  0.727 &  0.767 &  0.747 &  0.778 &  2.877 &  2.865 &  2.726 &  2.708 \\
					&ReM &  0.351 &  0.416 &  0.346 &  0.419 &  0.687 &  0.762 &  0.748 &  0.835 &  2.710 &  2.689 &  2.495 &  2.914 \\
					\multirow{2}{*}{Bayes} &CRE    &  0.360 &  0.385 &  0.334 &  0.371 &  0.671 &  0.675 &  0.628 &  0.680 &  1.120 &  1.674 &  1.281 &  1.361 \\
					&ReM     &  0.340 &  0.404 &  0.330 &  0.420 &  0.669 &  0.673 &  0.631 &  0.737 &  1.067 &  1.676 &  1.199 &  1.265 \\
					\hline\hline
				\end{tabular}
			\end{center}
		\end{scriptsize}
	\end{table}
\end{landscape}
	
\begin{landscape}
	\begin{table}[tbp]
		\begin{scriptsize}
			\caption{Coverage rate of 95\% intervals for different methods under DGP2, measured in percentage differences from the nominal level 0.95, $100\times (\text{CRate}_m-0.95)$.}
			\label{T:CR:DGP2}
			\begin{center}
				\begin{tabular}{llS[table-format=1.3]S[table-format=1.3]S[table-format=1.3]S[table-format=1.3]S[table-format=1.3]S[table-format=1.3]S[table-format=1.3]S[table-format=1.3]S[table-format=1.3]S[table-format=1.3]S[table-format=1.3]S[table-format=1.3]}
					\hline\hline
					& & \multicolumn{4}{c}{$p_{co}=0.85$} & \multicolumn{4}{c}{$p_{co}=0.5$} & \multicolumn{4}{c}{$p_{co}=0.15$}\\
					method & rand & {case 1} & {case 2} & {case 3} & {case 4} & {case 1} & {case 2} & {case 3} & {case 4} & {case 1} & {case 2} & {case 3} & {case 4}\\
					\hline
					& & \multicolumn{12}{c}{$K=5,\ n=200$}\\
					\multirow{2}{*}{Wald} &CRE   &  2.3 &  1.8 &  2.0 &   1.0 &  1.7 &  1.3 &  0.8 &   1.0 &  2.6 &  2.5 &  2.3 &  3.1 \\
					&ReM      &  2.8 &  1.4 &  3.5 &  -0.9 &  1.3 &  0.9 &  0.2 &  -1.5 &  3.7 &  3.7 &  3.7 &  3.7 \\
					\multirow{2}{*}{Reg-EHW} &CRE &  2.9 &  1.4 &  2.5 &   1.1 &  1.6 &  0.8 &  1.7 &   0.9 &  3.6 &  3.3 &  3.8 &  4.0 \\
					&ReM &  3.3 &  1.4 &  3.2 &   0.6 &  1.5 &  1.3 &  1.0 &   0.0 &  3.9 &  4.1 &  4.1 &  3.6 \\
					\multirow{2}{*}{Reg-HC2} &CRE &  3.2 &  1.9 &  3.0 &   1.9 &  2.3 &  1.5 &  2.5 &   1.3 &  3.8 &  3.9 &  4.0 &  4.3 \\
					&ReM &  3.6 &  1.8 &  3.5 &   1.2 &  2.0 &  1.5 &  1.6 &   0.7 &  4.2 &  4.4 &  4.2 &  3.9 \\
					\multirow{2}{*}{Reg-HC3} &CRE &  3.5 &  2.2 &  3.4 &   2.5 &  2.7 &  2.1 &  3.1 &   1.7 &  3.8 &  4.0 &  4.3 &  4.7 \\
					&ReM &  3.9 &  2.4 &  4.0 &   1.9 &  2.7 &  2.0 &  2.6 &   1.4 &  4.5 &  4.5 &  4.3 &  4.4 \\
					\multirow{2}{*}{Bayes} &CRE    &  2.1 &  0.6 &  0.7 &   0.3 &  1.6 &  0.6 &  1.7 &   2.1 &  3.7 &  4.3 &  4.1 &  4.8 \\
					&ReM     &  2.1 &  0.2 &  0.8 &   0.4 &  1.6 &  0.7 &  2.5 &   1.8 &  3.3 &  4.0 &  4.0 &  4.5 \\
					\hline
					& &\multicolumn{12}{c}{$K=5,\ n=400$}\\
					\multirow{2}{*}{Wald} &CRE     &  2.3 &  1.6 &  1.9 &  0.7 &  0.8 &  1.0 &  1.4 &  0.8 &  1.8 &  2.5 &  3.0 &  0.8 \\
					&ReM      &  3.1 &  0.9 &  2.5 &  2.3 &  1.8 &  0.3 &  1.5 &  1.2 &  3.0 &  3.0 &  3.1 &  2.6 \\
					\multirow{2}{*}{Reg-EHW} &CRE &  3.7 &  2.2 &  3.4 &  1.8 &  2.4 &  1.7 &  2.1 &  1.8 &  3.2 &  2.9 &  3.9 &  2.8 \\
					&ReM&  3.7 &  1.8 &  2.7 &  2.6 &  1.9 &  0.8 &  2.0 &  1.6 &  3.4 &  3.1 &  3.2 &  2.9 \\
					\multirow{2}{*}{Reg-HC2} &CRE &  3.7 &  2.6 &  3.5 &  2.0 &  2.6 &  1.9 &  2.6 &  2.1 &  3.4 &  3.0 &  4.1 &  3.3 \\
					&ReM &  3.8 &  2.1 &  2.9 &  2.7 &  1.9 &  1.1 &  2.1 &  1.7 &  3.4 &  3.1 &  3.3 &  3.2 \\
					\multirow{2}{*}{Reg-HC3} &CRE &  3.7 &  2.9 &  3.6 &  2.1 &  2.6 &  2.0 &  2.9 &  2.3 &  3.6 &  3.1 &  4.2 &  3.6 \\
					&ReM&  3.9 &  2.3 &  3.3 &  2.9 &  2.3 &  1.3 &  2.4 &  1.7 &  3.8 &  3.2 &  3.3 &  3.4 \\
					\multirow{2}{*}{Bayes} &CRE    &  3.2 &  1.3 &  3.1 &  1.7 &  2.0 &  1.6 &  2.6 &  2.2 &  2.6 &  4.4 &  3.8 &  4.0 \\
					&ReM     &  3.2 &  0.9 &  2.9 &  2.6 &  2.2 &  1.7 &  2.7 &  2.1 &  2.9 &  4.7 &  3.4 &  4.2 \\
					\hline
					& &\multicolumn{12}{c}{$K=10,\ n=200$}\\
					\multirow{2}{*}{Wald} &CRE      &  2.0 &   1.5 &  2.5 &  0.3 &  0.5 &   0.2 &  1.1 &  0.2 &  0.5 &  0.7 &   2.9 &  2.1 \\
					&ReM    &  3.2 &   1.8 &  3.0 &  2.6 &  1.9 &   1.3 &  1.4 &  2.2 &  2.9 &  2.9 &   4.0 &  3.6 \\
					\multirow{2}{*}{Reg-EHW} &CRE&  2.7 &   0.8 &  3.2 &  0.8 &  0.6 &  -0.1 &  1.2 &  0.3 &  3.3 &  3.3 &   3.8 &  3.6 \\
					&ReM &  2.2 &  -0.5 &  2.5 &  1.8 &  0.4 &  -1.3 &  0.2 &  1.1 &  2.9 &  2.9 &   3.6 &  4.0 \\
					\multirow{2}{*}{Reg-HC2} &CRE &  3.2 &   1.8 &  3.7 &  1.8 &  1.9 &   1.5 &  1.9 &  1.4 &  4.0 &  4.0 &   4.3 &  4.1 \\
					&ReM&  3.0 &   0.9 &  3.2 &  3.0 &  1.9 &   0.5 &  1.3 &  2.5 &  3.4 &  3.7 &   3.8 &  4.4 \\
					\multirow{2}{*}{Reg-HC3} &CRE &  3.9 &   3.2 &  4.4 &  3.4 &  3.2 &   2.6 &  2.7 &  2.2 &  4.5 &  4.2 &   4.8 &  4.4 \\
					&ReM &  4.1 &   2.6 &  3.7 &  3.9 &  2.4 &   2.0 &  2.4 &  3.3 &  4.1 &  4.0 &   4.6 &  4.9 \\
					\multirow{2}{*}{Bayes} &CRE    &  1.9 &   0.5 &  3.4 &  0.1 &  1.7 &   1.5 &  3.0 &  0.7 &  3.8 &  2.1 &  -4.8 &  3.3 \\
					&ReM     &  1.3 &  -0.6 &  2.9 &  1.1 &  1.3 &   0.7 &  2.3 &  1.3 &  2.8 &  1.3 &  -5.6 &  4.1 \\
					\hline
					& &\multicolumn{12}{c}{$K=10,\ n=400$}\\
										\multirow{2}{*}{Wald} &CRE     &  1.9 &  0.7 &  1.9 &  1.5 &  0.8 &  0.6 &  1.0 &  0.8 &  2.0 &  2.4 &  1.7 &  2.7 \\
					&ReM    &  3.3 &  2.3 &  3.7 &  2.2 &  2.3 &  1.7 &  2.8 &  1.6 &  3.8 &  3.9 &  3.4 &  3.4 \\
					\multirow{2}{*}{Reg-EHW} &CRE &  2.9 &  1.6 &  3.2 &  2.7 &  1.3 &  0.3 &  1.4 &  0.7 &  2.8 &  2.6 &  2.9 &  3.7 \\
					&ReM &  2.6 &  1.8 &  2.7 &  1.2 &  1.5 &  0.8 &  1.5 &  0.9 &  2.5 &  2.9 &  2.8 &  2.6 \\
					\multirow{2}{*}{Reg-HC2} &CRE &  3.4 &  1.9 &  3.4 &  3.2 &  1.6 &  0.7 &  1.7 &  1.9 &  3.0 &  3.3 &  3.0 &  3.9 \\
					&ReM &  2.9 &  1.9 &  3.2 &  1.6 &  1.8 &  1.3 &  1.8 &  1.6 &  3.2 &  3.4 &  3.3 &  3.2 \\
					\multirow{2}{*}{Reg-HC3} &CRE &  3.7 &  2.0 &  3.6 &  3.5 &  2.2 &  1.9 &  2.4 &  2.4 &  3.3 &  3.5 &  3.3 &  4.3 \\
					&ReM &  3.3 &  2.3 &  3.6 &  2.1 &  2.4 &  1.8 &  1.9 &  2.0 &  3.7 &  3.8 &  3.7 &  3.8 \\
					\multirow{2}{*}{Bayes} &CRE     &  2.7 &  0.6 &  2.8 &  2.3 &  1.1 &  1.0 &  1.8 &  1.9 &  3.9 &  1.5 &  4.0 &  3.6 \\
					&ReM    &  2.8 &  0.9 &  3.3 &  1.1 &  1.3 &  1.0 &  2.2 &  1.1 &  4.1 &  2.2 &  4.4 &  3.4 \\
					\hline					\hline
				\end{tabular}
			\end{center}
		\end{scriptsize}
	\end{table}
\end{landscape}	

\begin{landscape}
	\begin{table}[tbp]
		\begin{scriptsize}
			\caption{Median length of 95\% interval for different methods under DGP2.}
			\label{T:LEN:DGP2}
			\begin{center}
				\begin{tabular}{llS[table-format=1.3]S[table-format=1.3]S[table-format=1.3]S[table-format=1.3]S[table-format=1.3]S[table-format=1.3]S[table-format=1.3]S[table-format=1.3]S[table-format=1.3]S[table-format=1.3]S[table-format=1.3]S[table-format=1.3]}
					\hline\hline
					& & \multicolumn{4}{c}{$p_{co}=0.85$} & \multicolumn{4}{c}{$p_{co}=0.5$} & \multicolumn{4}{c}{$p_{co}=0.15$}\\
					method & rand &{case 1} & {case 2} & {case 3} & {case 4} & {case 1} & {case 2} & {case 3} & {case 4} & {case 1} & {case 2} & {case 3} & {case 4}\\
					\hline
					& &\multicolumn{12}{c}{$K=5,\ n=200$}\\
					\multirow{2}{*}{Wald} &CRE      &  3.607 &  3.669 &  3.616 &  3.556 &  6.337 &  6.126 &  6.273 &  6.261 &  24.541 &  21.801 &  24.149 &  21.830 \\
					&ReM     &  2.924 &  2.792 &  2.917 &  2.533 &  5.163 &  4.658 &  5.120 &  4.518 &  19.892 &  17.168 &  19.771 &  15.893 \\
					\multirow{2}{*}{Reg-EHW} &CRE  &  2.797 &  2.815 &  2.813 &  2.644 &  5.246 &  5.165 &  5.279 &  4.787 &  18.321 &  18.263 &  18.651 &  16.436 \\
					&ReM &  2.754 &  2.771 &  2.782 &  2.616 &  5.197 &  5.122 &  5.257 &  4.714 &  18.597 &  18.467 &  18.569 &  16.618 \\
					\multirow{2}{*}{Reg-HC2} &CRE &  2.899 &  2.917 &  2.916 &  2.735 &  5.443 &  5.360 &  5.486 &  4.953 &  18.953 &  18.925 &  19.350 &  16.985 \\
					&ReM &  2.852 &  2.872 &  2.883 &  2.709 &  5.379 &  5.305 &  5.462 &  4.876 &  19.232 &  19.100 &  19.261 &  17.171 \\
					\multirow{2}{*}{Reg-HC3} &CRE &  3.004 &  3.023 &  3.026 &  2.835 &  5.646 &  5.565 &  5.707 &  5.131 &  19.618 &  19.609 &  20.077 &  17.564 \\
					&ReM &  2.955 &  2.977 &  2.989 &  2.805 &  5.583 &  5.502 &  5.678 &  5.046 &  19.909 &  19.758 &  19.985 &  17.763 \\
					\multirow{2}{*}{Bayes} &CRE       &  2.363 &  2.215 &  2.280 &  2.040 &  4.117 &  3.521 &  4.194 &  3.119 &  11.041 &   7.826 &  11.044 &   6.681 \\
					&ReM   &  2.328 &  2.183 &  2.240 &  2.011 &  4.087 &  3.512 &  4.167 &  3.093 &  10.778 &   7.795 &  10.813 &   6.735 \\
					\hline
					& &\multicolumn{12}{c}{$K=5,\ n=400$}\\
					\multirow{2}{*}{Wald} &CRE        &  2.533 &  2.502 &  2.546 &  2.529 &  4.509 &  4.434 &  4.402 &  4.505 &  15.588 &  15.119 &  15.373 &  15.992 \\
					&ReM    &  1.894 &  1.896 &  1.890 &  1.820 &  3.392 &  3.379 &  3.338 &  3.271 &  11.774 &  11.515 &  11.525 &  11.518 \\
					\multirow{2}{*}{Reg-EHW} &CRE   &  1.891 &  1.893 &  1.892 &  1.819 &  3.395 &  3.377 &  3.345 &  3.277 &  11.902 &  11.624 &  11.712 &  11.418 \\
					&ReM &  1.881 &  1.885 &  1.883 &  1.803 &  3.367 &  3.352 &  3.323 &  3.240 &  11.780 &  11.562 &  11.502 &  11.328 \\
					\multirow{2}{*}{Reg-HC2} &CRE &  1.925 &  1.927 &  1.925 &  1.850 &  3.457 &  3.439 &  3.405 &  3.332 &  12.102 &  11.836 &  11.915 &  11.606 \\
					&ReM &  1.915 &  1.919 &  1.916 &  1.834 &  3.427 &  3.411 &  3.382 &  3.295 &  11.989 &  11.767 &  11.703 &  11.512 \\
					\multirow{2}{*}{Reg-HC3} &CRE &  1.960 &  1.963 &  1.960 &  1.883 &  3.519 &  3.500 &  3.466 &  3.389 &  12.314 &  12.056 &  12.125 &  11.798 \\
					&ReM &  1.950 &  1.955 &  1.950 &  1.865 &  3.488 &  3.473 &  3.442 &  3.351 &  12.205 &  11.980 &  11.903 &  11.704 \\
					\multirow{2}{*}{Bayes} &CRE    &  1.799 &  1.800 &  1.785 &  1.721 &  3.073 &  3.116 &  2.997 &  2.834 &   6.813 &   6.773 &   7.569 &   5.402 \\
					&ReM   &  1.795 &  1.794 &  1.777 &  1.711 &  3.048 &  3.089 &  2.967 &  2.803 &   6.826 &   6.770 &   7.573 &   5.395 \\
					\hline
					& &\multicolumn{12}{c}{$K=10,\ n=200$}\\
					\multirow{2}{*}{Wald} &CRE        &  5.092 &  5.137 &  5.224 &  5.384 &  9.332 &  9.326 &  9.522 &  9.760 &  32.799 &  33.307 &  32.442 &  35.349 \\
					&ReM   &  4.197 &  4.286 &  4.256 &  4.519 &  7.681 &  7.809 &  7.866 &  8.228 &  27.475 &  28.283 &  27.946 &  29.986 \\
					\multirow{2}{*}{Reg-EHW} &CRE  &  3.668 &  3.822 &  3.719 &  4.080 &  6.585 &  6.778 &  6.708 &  7.350 &  23.284 &  24.550 &  23.836 &  26.886 \\
					&ReM &  3.620 &  3.759 &  3.649 &  3.990 &  6.506 &  6.724 &  6.611 &  7.248 &  23.233 &  24.608 &  23.879 &  26.236 \\
					\multirow{2}{*}{Reg-HC2} &CRE &  3.911 &  4.079 &  3.951 &  4.357 &  7.002 &  7.227 &  7.108 &  7.858 &  24.745 &  26.176 &  25.253 &  28.641 \\
					&ReM &  3.855 &  4.010 &  3.875 &  4.256 &  6.922 &  7.156 &  7.012 &  7.739 &  24.676 &  26.172 &  25.330 &  28.067 \\
					\multirow{2}{*}{Reg-HC3} &CRE &  4.176 &  4.356 &  4.204 &  4.657 &  7.471 &  7.714 &  7.558 &  8.404 &  26.378 &  27.901 &  26.784 &  30.644 \\
					&ReM &  4.115 &  4.284 &  4.122 &  4.547 &  7.375 &  7.647 &  7.440 &  8.281 &  26.218 &  27.859 &  26.898 &  30.040 \\
					\multirow{2}{*}{Bayes} &CRE       &  3.461 &  3.638 &  3.734 &  3.881 &  6.209 &  6.642 &  5.853 &  6.744 &  10.740 &  11.329 &  11.772 &  13.817 \\
					&ReM    &  3.409 &  3.589 &  3.656 &  3.802 &  6.168 &  6.550 &  5.764 &  6.621 &  10.604 &  11.209 &  11.726 &  13.751 \\
					\hline
					& &\multicolumn{12}{c}{$K=10,\ n=400$}\\
					\multirow{2}{*}{Wald} &CRE        &  3.477 &  3.602 &  3.467 &  3.614 &  6.051 &  6.262 &  6.124 &  6.473 &  20.973 &  21.455 &  20.988 &  22.234 \\
					&ReM     &  2.939 &  3.028 &  2.922 &  3.065 &  5.151 &  5.264 &  5.194 &  5.554 &  18.122 &  18.270 &  17.783 &  19.456 \\
					\multirow{2}{*}{Reg-EHW} &CRE   &  2.581 &  2.639 &  2.551 &  2.703 &  4.554 &  4.612 &  4.545 &  4.883 &  16.180 &  16.095 &  15.642 &  16.899 \\
					&ReM &  2.546 &  2.604 &  2.521 &  2.679 &  4.490 &  4.536 &  4.509 &  4.854 &  16.057 &  16.033 &  15.419 &  16.983 \\
					\multirow{2}{*}{Reg-HC2} &CRE &  2.655 &  2.713 &  2.623 &  2.786 &  4.687 &  4.744 &  4.674 &  5.036 &  16.637 &  16.554 &  16.081 &  17.414 \\
					&ReM &  2.619 &  2.679 &  2.591 &  2.761 &  4.618 &  4.667 &  4.637 &  5.005 &  16.505 &  16.480 &  15.861 &  17.494 \\
					\multirow{2}{*}{Reg-HC3} &CRE &  2.731 &  2.791 &  2.698 &  2.873 &  4.821 &  4.882 &  4.810 &  5.195 &  17.111 &  17.034 &  16.546 &  17.958 \\
					&ReM &  2.695 &  2.756 &  2.664 &  2.847 &  4.752 &  4.799 &  4.770 &  5.164 &  16.976 &  16.943 &  16.319 &  18.025 \\
					\multirow{2}{*}{Bayes} &CRE      &  2.435 &  2.459 &  2.436 &  2.601 &  4.269 &  4.236 &  4.186 &  4.398 &   9.619 &   9.037 &   9.489 &  10.593 \\
					&ReM &  2.413 &  2.436 &  2.407 &  2.577 &  4.245 &  4.194 &  4.165 &  4.382 &   9.617 &   9.015 &   9.440 &  10.435 \\
					\hline		\hline
				\end{tabular}
			\end{center}
		\end{scriptsize}
	\end{table}
\end{landscape}

We first examine the performance under the linear data generation process DGP1.  In terms of median absolute error, under either CRE or ReM, for almost all of the settings, the Wald method has the largest median absolute error and the Bayesian method has the smallest median absolute error.  The improvement by the Bayesian method is especially large when the fraction of compliers is small ($p_{co}=0.15$).  If we compare CRE and ReM, for the Wald method, rerandomization yields smaller median absolute error in all of the settings, and the improvement due to rerandomization can be rather large in some settings.  For the regression adjustment method and the Bayesian method, rerandomization yields smaller median absolute error in most of the settings, but yields larger median absolute error in some settings.  The difference between CRE and ReM is rather small.  When $K=10$, the regression adjustment method under CRE has smaller median absolute error than the Wald method under ReM.

In terms of coverage rate, out of 96 settings, Reg-EHW has slight under-coverage in 2 settings, with the largest value of under coverage being 0.6\%.  The Bayesian method has some under-coverage in 7 settings, with the largest value of under coverage being 2.9\%.  The four settings in which the Bayesian method has an under coverage rate of more than 2\% all have $p_{co}=0.15$.  The Wald, Reg-HC2 and Reg-HC3 methods do not have under coverage.

In terms of median length of 95\% intervals, the Bayesian method has smaller median interval length than the Wald method and the regression adjustment methods.   The improvement by the Bayesian method is especially large when the fraction of compliers is small ($p_{co}=0.15$).  Among the three regression adjustment methods, Reg-EHW has the smallest median interval length, and Reg-HC3 has the largest median interval length.  Under CRE, the Wald method has larger median interval length than Reg-HC3.  Under ReM, the relative performance of the Wald method and the regression adjustment methods is not clear-cut.  When $K=5$, in almost all of the settings, the Wald method has larger median interval length than Reg-EHW but smaller median interval length than Reg-HC2; when $K=10$, the Wald method has larger median interval length than Reg-HC3. If we compare CRE and ReM, for all of the methods and in almost all of the settings, rerandomization yields smaller median interval length.  For the Wald method, the improvement due to rerandomization is rather large in some settings.  For the regression adjustment method and the Bayesian method, the difference between CRE and ReM is rather small.  When $K=10$, Reg-EHW under CRE has smaller median interval length than the Wald method under ReM.

We now examine the performance under the nonlinear data generation process DGP2.  In general, the results are quite similar to those for DGP1, and therefore we only discuss results that are different from those for DGP1.  In terms of coverage rate, out of 96 settings, the Wald method has some under coverage in 2 settings, with the largest under coverage being 1.5\%.  The Reg-EHW method has some under coverage in 3 settings, with the largest under coverage being 1.3\%.  The Bayesian method has some under coverage in 3 settings, with the largest under coverage being 5.6\%. The two settings in which the Bayesian method has an under coverage rate of more than 2\% both have $p_{co}=0.15$.

In terms of median length of 95\% intervals, under ReM when $K=5$, the median interval length of the Wald method is smaller than that of the Reg-EHW for some settings, larger than that of Reg-EHW but smaller than that of Reg-HC2 for some settings, and larger than that of Reg-HC2 and smaller than that of Reg-HC3 for other settings.  If we compare CRE and ReM, for the regression adjustment methods, rerandomization yields smaller median interval length in most of the settings, but yields larger median interval length in some settings.

To summarize, the Bayesian method performs the best because it is stable, it yields smallest median absolute error and smallest median interval length, regardless of whether CRE or ReM is adopted.  The improvement by the Bayesian method is especially large when the fraction of compliers is small.  A caveat is that when the fraction of compliers is small, the Bayesian method can have some under coverage in some settings.  With CRE, the Wald method has the largest median absolute error and largest median interval length.  This can be improved to a large extent by regression adjustment and/or rerandomization.  In terms of the three regression adjustment methods, Reg-EHW performs the best since it yields smallest median interval length.  When the number of covariates is large, Reg-EHW under CRE works better than the Wald method under ReM.  With Reg-EHW or the Bayesian method, the difference in performance between CRE or ReM is rather small.  Given rerandomization, Reg-EHW has smaller median absolute error than the Wald method, and Reg-EHW also has smaller median interval length than the Wald method when the number of covariates is large.

\section{Application to a Job Training Experiment with Non-compliance}
The Job Search Intervention Study (JOBS II) dataset comes from a field experiment designed and conducted by \cite{vinokur1995impact} that investigates the efficacy of a job training intervention on unemployed workers. Participants were randomly selected to attend the JOBS II training program that taught job-search skills and coping strategies for dealing with setbacks in the job-search process.  In the dataset, 372 participants were selected to attend the training program and actually attended, 228 participants were selected to attend the training program but did not attend, and 299 participants were not selected to attend the training program and did not attend.  One outcome of interest is a continuous variable measuring the level of job-search self-efficacy with values from 1 to 5.  The covariates include age, gender, ethnicity, marital status, monthly income and educational attainment.  This data set has been analyzed in Chapter 21 of \cite{ding2023first} using the method with four possible forms of confidence sets in \cite{li2017general}.

For different methods presented in this paper under CRE, Table~\ref{real_data} present point estimates of $p_{co}^{samp}$, and point estimates and 95\% intervals of $\tau_{CACE}^{samp}$.  The estimated fraction of compliers is around 0.62 for all methods.  The points estimates of $\tau_{CACE}^{samp}$ from different methods are similar.  The Bayesian method has the shortest 95\% interval for $\tau_{CACE}^{samp}$. The Reg-EHW method has slightly shorter 95\% interval than the Wald method, but the Reg-HC2 and Reg-HC3 methods have longer 95\% intervals than the Wald method.
	
	\begin{table}[htbp]
		\centering
		\caption{Point estimates of $p_{co}^{samp}$, and point estimates and 95\% intervals of $\tau_{CACE}^{samp}$ from different methods for the JOBS II dataset}
		\label{real_data}
		\begin{tabular}{lccc}
			\hline\hline
            & point estimate & point estimate & 95\% interval\\
			method  & of $p_{co}^{samp}$  & of $\tau_{CACE}^{samp}$ &    of $\tau_{CACE}^{samp}$\\
			\hline
			Wald  & 0.620 & 0.109 &  [-0.050, 0.268] \\
			Reg-EHW &  0.618 & 0.118 &  [-0.039, 0.274] \\
			Reg-HC2 & 0.618 & 0.118 &  [-0.043, 0.278] \\
			Reg-HC3 & 0.618 & 0.118 &  [-0.046, 0.282] \\
			Bayes  & 0.616 &0.110	& [-0.012, 0.236]\\
					\hline		\hline
		\end{tabular}
	\end{table}
	
\section{Discussion}
In pragmatic randomized control trials, incomplete adherence and/or incomplete compliance to the assigned treatment sequences is common. This paper focuses on inference about the sample complier average causal effect.  We discuss three inference strategies: the Wald estimator, regression adjustment estimators and model-based Bayesian inference.  We compare their small sample performance using a Monte Carlo simulation, under both complete randomization and Mahalanobis distance based rerandomization.

The results from the Monte Carlo simulation shows that under either design, the Bayesian method performs the best because it is stable, it yields smallest median absolute error and smallest median interval length.  The improvement by the Bayesian method is especially large when the fraction of compliers is small.  Results also show that rerandomization can bring significant benefits for the Wald method, but makes little difference when the regression adjustment method or the Bayesian method is used.

The asymptotic results for the Wald estimator and the regression adjustment estimators are contingent on the Mahalanobis distance based rerandomization procedure, but the Bayesian method is not contingent on this procedure. This means that it is straightforward to use model-based Bayesian inference for any computer assisted experimental designs that uses only the covariates, but this is not the case for the Wald method or the regression adjustment method.

\section*{Appendix}
\subsection*{A. Proof of Lemmas, Theorems and Propositions}
\renewcommand{\appendixname}{Appendix~\Alph{section}}
\setcounter{theorem}{0}
\renewcommand\thetheorem{A\arabic{theorem}}
Appendix A contains all the propositions and proofs omitted in the main article.

\begin{proposition}
	The interval estimator \eqref{ci_Wald} under CRE equals to that obtained by the delta method.
\end{proposition}
\subsubsection*{Proof of Proposition A1:}
For $z=0,1$, define
\begin{equation}
	\begin{aligned}
		S_{Y_z}^2&=\frac{1}{n_z-1}\sum_{i:Z_i=z}\left(Y^{obs}_z-\overline{Y}^{obs}_z\right)^2,\\ S_{W_z}^2&=\frac{1}{n_z-1}\sum_{i:Z_i=z}\left(W^{obs}_z-\overline{W}^{obs}_z\right)^2,\\
		S_{Y_z,W_z}&=\frac{1}{n_z-1}\sum_{i:Z_i=z}\left(Y^{obs}_z-\overline{Y}^{obs}_z\right)\left(W^{obs}_z-\overline{W}^{obs}_z\right).
	\end{aligned}
\end{equation}
Expanding the expression of $\widehat{Var}(\widetilde{\tau}_A)$, we have
\begin{equation}
	\label{delta_method}
	\begin{aligned}
		\widehat{Var}(\widetilde{\tau}_A)=&\widehat{Var}(\widehat{ITT}_Y)+\frac{\widehat{ITT}_{Y}^2}{\widehat{ITT}_{W}^2}\cdot\widehat{Var}(\widehat{ITT}_W)\\&-2\cdot\frac{\widehat{ITT}_{Y}}{\widehat{ITT}_{W}}\cdot\widehat{Var}(\widehat{ITT}_Y,\widehat{ITT}_W),
	\end{aligned}
\end{equation}
where $\widehat{Var}(\widehat{ITT}_Y)$ and $\widehat{Var}(\widehat{ITT}_W)$ are standard variance estimators
\begin{equation}
	\widehat{Var}(\widehat{ITT}_Y)=\frac{S_{Y_1}^2}{n_1}+\frac{S_{Y_0}^2}{n_0},\quad\widehat{Var}(\widehat{ITT}_W)=\frac{S_{W_1}^2}{n_1}+\frac{S_{W_0}^2}{n_0}
\end{equation}
and $\widehat{Var}(\widehat{ITT}_Y,\widehat{ITT}_W)$ is the standard covariance estimator
\begin{equation}
	\widehat{Var}(\widehat{ITT}_Y,\widehat{ITT}_W)=\frac{S_{Y_1,W_1}}{n_1}+\frac{S_{Y_0,W_0}}{n_0}.
\end{equation}
Substituting (\ref{delta_method}) into (\ref{ci_Wald}). we find our proposed confidence interval equals to the super-population confidence interval obtained by the delta method (see e.g. \cite{imbens2015causal}, Ch. 23).

\begin{proposition}
	For general outcome variable $O_i$, the estimated coefficient of $Z_i$ in the OLS regression of $O_i$ on $(1, Z_i,\bm{x}_i^{*},Z_i\bm{x}_i^{*})$ equals to $\widehat{ITT}_O^{adj}$ given by
	\begin{equation}
		\label{ittw_adjust_general}
		\widehat{ITT}_O^{adj}=\frac{1}{n_1}\sum_{i:\ Z_i=1}(O_i(1)-\widehat{\boldsymbol{\beta}}_{O,1}\boldsymbol{x}_i^*)-\frac{1}{n_0}\sum_{i:\ Z_i=0}(O_i(0)-\widehat{\boldsymbol{\beta}}_{O,0}\boldsymbol{x}_i^*),
	\end{equation}
	where $\widehat{\boldsymbol{\beta}}_{O,z}$ is the estimated coefficient vector in the OLS regression of $O_i$ on $(1,\bm{x}_i^{*})$ based on un units $Z_i=z$.
\end{proposition}
\subsubsection*{Proof of Proposition A2:}
In the regression of $O_i$ on $(1,Z_i,\bm{x}_i^{*},Z_i\bm{x}_i^{*})$, let the estimated intercept be $\widehat{\beta}_0$ and the estimated coefficient of $Z_i$ be $\widehat{\beta}_Z$. Also let the estimated coefficient vector of $\bm{x}_i^{*}$ and $Z_i\bm{x}_i^{*}$ be $\widehat{\bm{\beta}}_{O,\bm{x}}$ and $\widehat{\bm{\beta}}_{O,Z\bm{x}}$. Denote $\varepsilon_{O,i}$ as the residual of unit $i$.
We could write the fully interact regression equation as
\begin{equation}
	O_i=\widehat{\beta}_0+\widehat{\beta}_ZZ_i+\widehat{\bm{\beta}}_{O,\bm{x}}\bm{x}_i^{*}+\widehat{\bm{\beta}}_{O,Z\bm{x}}(Z_i\bm{x}_i^{*})+\varepsilon_{O,i}.
\end{equation}
By orthogonality of regressor $Z_i$ with the residual, we have
\begin{equation}
	\label{orthogonality_eliminate}
	\widehat{\beta}_Z=\frac{1}{n_1}\sum_{i:\ Z_i=1}(O_i(1)-\widehat{\bm{\beta}}_{O,\bm{x}}\boldsymbol{x}_i^*-\widehat{\bm{\beta}}_{O,Z\bm{x}}\boldsymbol{x}_i^*)-\frac{1}{n_0}\sum_{i:\ Z_i=0}(O_i(0)-\widehat{\bm{\beta}}_{O,\bm{x}}\boldsymbol{x}_i^*)
\end{equation}

As the number of parameters in the OLS model of $O_i^{obs}$ on $(1,Z_i,\bm{x}_i^{*},Z_i\bm{x}_i^{*})$ based on all units equal to the total number of parameters in the OLS models of $O_i^{obs}$ on $(1,\bm{x}_i^{*})$ based on units with $Z_i=z$, $z=0,1$, we have the following correspondence:
\begin{equation}
	\label{correspondence}
	\begin{aligned}
		\widehat{\bm{\beta}}_{O,0}&=\widehat{\bm{\beta}}_{O,\bm{x}};\\
		\widehat{\bm{\beta}}_{O,1}&=\widehat{\bm{\beta}}_{O,\bm{x}}+\widehat{\bm{\beta}}_{O,Z\bm{x}}.
	\end{aligned}
\end{equation}
Substituting \eqref{correspondence} in \eqref{orthogonality_eliminate} completes the proof.

\subsubsection*{Proof of Theorem 1:}

For $z=0,1$, let $\mathbb{S}_{W_z}$ denote the finite population variance of $W_i(z)$, which are bounded values since $W_i(z)$'s are binary. Noting that $E(\widehat{ITT}_W)=p_{co}^{samp}$ and $Var(\widehat{ITT}_W)\leqslant\mathbb{S}^2_{W_1}/n_1+\mathbb{S}^2_{W_0}/n_0$, by Chebyshev's inequlity
\begin{align*}
	&\lim_{n\rightarrow\infty}Pr\left(|\widehat{ITT}_W-p_{co}^{samp}|\geqslant\epsilon\right)\\
	\leqslant&\lim_{n\rightarrow\infty}\frac{Var(\widehat{ITT}_W)}{\epsilon^2}\leqslant\lim_{n\rightarrow\infty}\frac{1}{\epsilon^2}\left(\frac{\mathbb{S}^2_{W_1}}{n_1}+\frac{\mathbb{S}^2_{W_0}}{n_0}\right)=0\quad\forall\epsilon>0,
\end{align*}
i.e., $\widehat{ITT}_W-p_{co}^{samp}=o_p(1)$. In a similar way, we have $\widehat{ITT}_Y-ITT_Y^{samp}=o_p(1)$, as we impose bounds on the sequences of $\mathbb{S}_{Y_z}^2$, $z=0,1$.

Note that
\begin{equation}
	\label{ratio_converge_in_p}
	\begin{aligned}
		&\left|\frac{\widehat{ITT}_Y}{\widehat{ITT}_W}-\frac{ITT_Y^{samp}}{p_{co}^{samp}}\right|\\
		\leqslant&\left|\frac{\widehat{ITT}_Y-ITT_Y^{samp}}{\widehat{ITT}_W}\right|+\left|\frac{p_{co}^{samp}-\widehat{ITT}_W}{\widehat{ITT}_W}\right|\left|\frac{ITT_Y^{samp}}{p_{co}^{samp}}\right|.
	\end{aligned}
\end{equation}
Since $p_{co}^{samp}$ has a positive limit inferior, we know that $\widehat{ITT}_W$ has a positive limit inferior in probability. By continuous mapping theorem, the first term in \eqref{ratio_converge_in_p} is $o_p(1)$. By assumption, we also know that $\left|ITT_Y^{samp}/p_{co}^{samp}\right|$ has a finite limit superior. By continuous mapping theorem, the second term in \eqref{ratio_converge_in_p} is $o_p(1)$. We conclude that $\widehat{\tau}^{Wald}_{CACE}-\tau_{CACE}^{samp}=o_p(1)$.

\subsubsection*{Proof of Proposition 1:}

We first define some intermediate quantities for $z=0,1$:
\begin{align*}
	\overline{A}^{grp}_z&=\sum_{i:Z_i=z}A_i(z)/n_z,\\
	S_{A_z}^2&=\frac{1}{n_z-1}\sum_{i:Z_i=z}\left(A_i(z)-\overline{A}^{grp}_z\right)^2,\\
	S_{A_z, W_z}&=\frac{1}{n_z-1}\sum_{i:Z_i=z}\left(A_i(z)-\overline{A}^{grp}_z\right)\left(W_i(z)-\overline{W}^{obs}_z\right).
\end{align*}
Also define
\begin{equation*}
	\widetilde{Var}(\widetilde{\tau}_A)=\frac{S_{A_1}^2}{n_1}+\frac{S_{A_0}^2}{n_0}.
\end{equation*}
Under condition \eqref{lindeberg-feller}, we can apply Proposition 1 in \cite{li2017general} to conclude that $\widetilde{Var}(\widetilde{\tau}_A)/Var(\widetilde{\tau}_A)^{+}\stackrel{p}{\longrightarrow}1$ as $n\rightarrow\infty$. We will next show that $|\widehat{Var}(\widetilde{\tau}_A)-\widetilde{Var}(\widetilde{\tau}_A)|/Var(\widetilde{\tau}_A)^{+}\stackrel{p}{\longrightarrow}0$ as $n\rightarrow\infty$, which implies that $\widehat{Var}(\widetilde{\tau}_A)/Var(\widetilde{\tau}_A)^{+}$ $\stackrel{p}{\longrightarrow}1$ as $n\rightarrow\infty$.

For $z=0,1$, noticing that $\widehat{A}_i(z)=A_i(z)+W_i(z)\left(\tau_{CACE}^{samp}-\widehat{\tau}_{CACE}^{Wald}\right)$, we can write $S_{\widehat{A}_z}^2$ as
\begin{equation}
\begin{aligned}
S_{\widehat{A}_z}^2=&\frac{1}{n_z-1}\sum_{i:Z_i=z}\left(\left(A_i(z)-\overline{A}^{grp}_z\right)+\left(\tau_{CACE}^{samp}-\widehat{\tau}_{CACE}^{Wald}\right)\left(W_i(z)-\overline{W}^{obs}_z\right)\right)^2\\
=&S_{A_z}^2+2\left(\tau_{CACE}^{samp}-\widehat{\tau}_{CACE}^{Wald}\right)S_{A_z,W_z}+\left(\tau_{CACE}^{samp}-\widehat{\tau}_{CACE}^{Wald}\right)^2S_{W_z}^2.
\end{aligned}
\end{equation}

The absolute difference between $\widehat{Var}(\widetilde{\tau}_A)$ and $\widetilde{Var}(\widetilde{\tau}_A)$ is then
		\begin{equation}
\begin{aligned}
			&\left|\widehat{Var}(\widetilde{\tau}_A)-\widetilde{Var}(\widetilde{\tau}_A)\right|\\
			=&\left|\frac{2\left(\tau_{CACE}^{samp}-\widehat{\tau}_{CACE}^{Wald}\right)S_{A_1,W_1}
+\left(\tau_{CACE}^{samp}-\widehat{\tau}_{CACE}^{Wald}\right)^2S^2_{W_1}}{n_1}\right.\\
			&\left.+\frac{2\left(\tau_{CACE}^{samp}-\widehat{\tau}_{CACE}^{Wald}\right)S_{A_0, W_0}
+\left(\tau_{CACE}^{samp}-\widehat{\tau}_{CACE}^{Wald}\right)^2S^2_{W_0}}{n_0}\right|\\
			\leqslant&\frac{2\left|\tau_{CACE}^{samp}-\widehat{\tau}_{CACE}^{Wald}\right|\left|S_{A_1, W_1}\right|+\left(\tau_{CACE}^{samp}-\widehat{\tau}_{CACE}^{Wald}\right)^2S^2_{W_1}}{n_1}\\
			&+\frac{2\left|\tau_{CACE}^{samp}-\widehat{\tau}_{CACE}^{Wald}\right|\left|S_{A_0,W_0}\right|+
\left(\tau_{CACE}^{samp}-\widehat{\tau}_{CACE}^{Wald}\right)^2S^2_{W_0}}{n_0}.
\end{aligned}\label{abs_diff}
\end{equation}

For $z=0,1$, because $W_i(z)$ is binary, we have
\begin{align}
S_{W_z}^2&\leqslant \frac{n_z}{4(n_z-1)},\label{S_Wz_leq}\\
|S_{A_z, W_z}|&\leqslant \frac{n_z}{n_z-1}\max_{1 \leqslant i \leqslant n}\left|A_{i}(z)-\overline{A}^{grp}_z\right|.\notag\\
&\leqslant \frac{n_z}{n_z-1}\left(\max_{1 \leqslant i \leqslant n}\left|A_{i}(z)-\overline{A}(z)\right|+|\widetilde{\tau}_A|\right)\label{S_AzWz_leq}
\end{align}
Plugging \eqref{S_Wz_leq} and \eqref{S_AzWz_leq} into \eqref{abs_diff}, we have
		\begin{equation}
\begin{aligned}
			&\left|\widehat{Var}(\widetilde{\tau}_A)-\widetilde{Var}(\widetilde{\tau}_A)\right|\\			\leqslant&\frac{\frac{2n_1}{n_1-1}\left|\tau_{CACE}^{samp}-\widehat{\tau}_{CACE}^{Wald}\right|\left(\max_{1 \leqslant i \leqslant n}|A_{i}(1)-\overline{A}(1)|+|\widetilde{\tau}_A|\right)}{n_1}+\frac{\left(\tau_{CACE}^{samp}-\widehat{\tau}_{CACE}^{Wald}\right)^2}{4(n_1-1)}\\
			&+\frac{\frac{2n_0}{n_0-1}\left|\tau_{CACE}^{samp}-\widehat{\tau}_{CACE}^{Wald}\right|\left(\max_{1 \leqslant i \leqslant n}\left|A_{i}(0)-\overline{A}(0)\right|+|\widetilde{\tau}_A|\right)}{n_0}+\frac{\left(\tau_{CACE}^{samp}-\widehat{\tau}_{CACE}^{Wald}\right)^2}{4(n_0-1)}.
		\end{aligned}\label{abs_diff2}
\end{equation}
		
Combining \eqref{est_Wald} and \eqref{tauA_tilde}, we have
		\begin{equation}
			\widehat{ITT}_{W}^{-1}\widetilde{\tau}_A=\widehat{\tau}_{CACE}^{Wald}-\tau_{CACE}^{samp}.\label{diff_Wald}
		\end{equation}
		Substituting \eqref{diff_Wald} into \eqref{abs_diff2} and dividing by $Var(\widetilde{\tau}_A)^{+}$, we have
\begin{equation}
		\begin{aligned}
&\left|\widehat{Var}(\widetilde{\tau}_A)-\widetilde{Var}(\widetilde{\tau}_A)\right|/Var(\widetilde{\tau}_A)^{+}\\
			=&\frac{\frac{2n_1}{n_1-1}\max_{1 \leqslant i \leqslant n}|A_{i}(1)-\overline{A}(1)|}{n_1\sqrt{Var(\widetilde{\tau}_A)^{+}}}\times|\widehat{ITT}_{W}^{-1}|\times\left|\frac{\widetilde{\tau}_A}{\sqrt{Var(\widetilde{\tau}_A)^{+}}}\right|\\
&+\left(\frac{2|\widehat{ITT}_{W}^{-1}|}{n_1-1}+\frac{\widehat{ITT}_{W}^{-2}}{4(n_1-1)}\right)\left(\frac{\widetilde{\tau}_A}{\sqrt{Var(\widetilde{\tau}_A)^{+}}}\right)^2\\
			&+\frac{\frac{2n_0}{n_0-1}\max_{1 \leqslant i \leqslant n}|A_{i}(0)-\overline{A}(0)|}{n_0\sqrt{Var(\widetilde{\tau}_A)^{+}}}\times|\widehat{ITT}_{W}^{-1}|\times\left|\frac{\widetilde{\tau}_A}{\sqrt{Var(\widetilde{\tau}_A)^{+}}}\right|\\
&+\left(\frac{2|\widehat{ITT}_{W}^{-1}|}{n_0-1}+\frac{\widehat{ITT}_{W}^{-2}}{4(n_0-1)}\right)\left(\frac{\widetilde{\tau}_A}{\sqrt{Var(\widetilde{\tau}_A)^{+}}}\right)^2\\
			\leqslant&\frac{4\max_{1 \leqslant i \leqslant n}|A_{i}(1)-\overline{A}(1)|}{\min(n_1, n_0)\sqrt{Var(\widetilde{\tau}_A)^{+}}}\times|\widehat{ITT}_{W}^{-1}|\times\left|\frac{\widetilde{\tau}_A}{\sqrt{Var(\widetilde{\tau}_A)^{+}}}\right|\\
			&+\frac{4\max_{1 \leqslant i \leqslant n}|A_{i}(0)-\overline{A}(0)|}{\min(n_1, n_0)\sqrt{Var(\widetilde{\tau}_A)^{+}}}\times|\widehat{ITT}_{W}^{-1}|\times\left|\frac{\widetilde{\tau}_A}{\sqrt{Var(\widetilde{\tau}_A)^{+}}}\right|\\
&+\frac{8|\widehat{ITT}_{W}^{-1}|+\widehat{ITT}_{W}^{-2}}{\min(n_1,n_0)}\left(\frac{\widetilde{\tau}_A}{\sqrt{Var(\widetilde{\tau}_A)^{+}}}\right)^2.
		\end{aligned}\label{abs_diff3}
\end{equation}
Condition (\ref{lindeberg-feller}) implies that
		\begin{equation}
			\lim_{n\rightarrow\infty}\frac{\max_{1 \leqslant i \leqslant n}\left|A_{i}(z)-\overline{A}(z)\right|}{\min(n_1, n_0)\cdot \sqrt{Var(\widetilde{\tau}_A)^{+}}}=0,\quad\text{for }z=0,1.
		\end{equation}
Because the limit inferior of $p_{co}^{samp}$ is positive (assumption (i)) and $\widehat{ITT}_{W}$ is a consistent estimator for $p_{co}^{samp}$, $\widehat{ITT}_{W}^{-1}$ has a positive limit superior in probability.  Assumption (ii) implies that $\min(n_1,n_0)\rightarrow \infty$ as $n\rightarrow\infty$.  We also know that $\widetilde{\tau}_A/\sqrt{Var(\widetilde{\tau}_A)}\stackrel{d}{\longrightarrow}N(0,1)$ as $n\rightarrow\infty$. Applying Slutsky's theorem, the right hand side of \eqref{abs_diff3} converges to 0 in probability.  Hence $|\widehat{Var}(\widetilde{\tau}_A)-\widetilde{Var}(\widetilde{\tau}_A)|/Var(\widetilde{\tau}_A)^{+}\stackrel{p}{\longrightarrow}0$ as $n\rightarrow \infty$.

\subsubsection*{Proof of Lemma 1:}

According to the definition of ReM, the probability of a random allocation being accepted, $p_a=n_a/n_A$, is a fixed number. If $\{U_n\}_{n=1}^{\infty}$, that is $o_p(n^k)$  for $k\in\mathbb{R}$ under CRE, then
\begin{align*}
	&\limsup_{n\rightarrow\infty}Pr(\left|U_n\right|\geqslant n^{k}\epsilon\mid\boldsymbol{Z}\in\mathcal{A}_a(\boldsymbol{X}))\\
	\leqslant&\limsup_{n\rightarrow\infty}\frac{1}{p_a}Pr(\left|U_n\right|\geqslant n^{k}\epsilon)\\
	=&\frac{1}{p_a}\lim_{n\rightarrow\infty}Pr(\left|U_n\right|\geqslant n^{k}\epsilon)\\
	=&0\quad\forall\epsilon>0,
\end{align*}
which means $\{U_n\}_{n=1}^{\infty}$ is $o_p(n^k)$ under ReM.

\subsubsection*{Proof of Proposition 2:}

According to Lemma \ref{dominated_converge_in_p}, we need only show that $\widehat{Var}(\widetilde{\tau}_A)_{\bm{x}}-Var\left(\widetilde{\tau}_A\right)^{+}_{\bm{x}}=o_p(n^{-1})$ and $\widehat{R}^2-R^{2-}=o_p(1)$ under CRE. In the rest of the proof, the underlying distribution becomes complete randomization.

By repeating the steps in the proof of Proposition \ref{conservative_CRE}, we obtain the following inequality
\begin{align}
	\label{bound_for_variance}
	&\left|S^2_{\widehat{A}_z}-S^2_{A_z}\right|/(n_zVar(\widetilde{\tau}_A)^{+})\notag\\
	\leqslant&\frac{4\max_{1 \leqslant i \leqslant n}|A_{i}(z)-\overline{A}(z)|}{\min(n_1, n_0)\sqrt{Var(\widetilde{\tau}_A)^{+}}}\times|\widehat{ITT}_{W}^{-1}|\times\left|\frac{\widetilde{\tau}_A}{\sqrt{Var(\widetilde{\tau}_A)^{+}}}\right|\\
	&+\frac{8|\widehat{ITT}_{W}^{-1}|+\widehat{ITT}_{W}^{-2}}{2\min(n_1,n_0)}\left(\frac{\widetilde{\tau}_A}{\sqrt{Var(\widetilde{\tau}_A)^{+}}}\right)^2.
\end{align}
for $z=0,1$. Under Condition \ref{strict_condition}, $n_zVar(\widetilde{\tau}_A)^{+}$ has a finite limit as $n\rightarrow\infty$. It has been already shown in the proof of Proposition \ref{conservative_CRE} that the right hand side of \eqref{bound_for_variance} is $o_p(1)$, which means $S^2_{\widehat{A}_z}-S^2_{A_z}=o_p(1)$ for $z=0,1$.

For $z=0,1$, Let $S_{W_z,\boldsymbol{x}}$ be the estimated covariance between $W_i(z)$ and $\boldsymbol{x}_i$; let the $k$th covariate of unit $i$ be $x_i^{(k)}$ and its group mean be $\overline{x}_z^{(k)}$. Using the Cauchy-Schwarz inequality for the $k$th component of $S_{W_z,\boldsymbol{x}}$, we have
\begin{align*}
	&\left(\frac{1}{n_z-1}\sum_{i:Z_{i}=z}(W_i(z)-\overline{W}^{obs}_z)(x_{i}^{(k)}-\overline{x}_z^{(k)})\right)^2\\
	\leqslant&\ S^2_{W_z}\frac{1}{n_z-1}\sum_{i:Z_{i}=z}(x_{i}^{(k)}-\overline{x}_z^{(k)})^2
\end{align*}
Also note that
\begin{equation*}
	\frac{1}{n_z-1}\sum_{i:Z_{i}=z}(x_{i}^{(k)}-\overline{x}_z^{(k)})^2\leqslant\frac{n-1}{n_z-1}\mathbb{S}_{\bm{x}\bm{x}}^{(k,k)},
\end{equation*}
where $\mathbb{S}_{\bm{x}\bm{x}}^{(k,k)}$ is the $k$th diagonal element of $\mathbb{S}_{\bm{x}\bm{x}}$, representing the finite population variance of the $k$th covariate.

For $z=0,1$, it follows that
\begin{align}
	\label{bound_for_covariance}
	&\left\Vert S_{\widehat{A}_z,\boldsymbol{x}}-S_{A_z,\boldsymbol{x}}\right\Vert_2/(n_zVar(\widetilde{\tau}_A)^{+})\notag\\
	=&\frac{\Vert S_{W_z,\boldsymbol{x}}\Vert_2|\widehat{\tau }_{CACE}^{\text{Wald}}-\tau_{CACE}^{\text{sample}}|}{n_zVar(\widetilde{\tau}_A)^{+}}\notag\\
	\leqslant&\frac{1}{\sqrt{n_z}}\frac{\sqrt{\frac{n-1}{n_z-1}\operatorname{trace}(\mathbb{S}_{\boldsymbol{x}\boldsymbol{x}})}}{\sqrt{n_zVar(\widetilde{\tau}_A)^{+}}}\times|\widehat{ITT}_{W}^{-1}|\times\left|\frac{\widetilde{\tau}_A}{\sqrt{Var(\widetilde{\tau}_A)^{+}}}\right|.
\end{align}
By assumption, $\frac{n-1}{n_z-1}\operatorname{trace}(\mathbb{S}_{\boldsymbol{x}\boldsymbol{x}})$ and $n_zVar(\widetilde{\tau}_A)^{+}$ have finite limits, meaning that the right hand side of \eqref{bound_for_covariance} is $o_p(1)$. And we conclude that $S_{\widehat{A}_z,\boldsymbol{x}}-S_{A_z,\boldsymbol{x}}=o_p(1)$ for $z=0,1$.

These results can be combined with Lemma A15 in \cite{li2018asymptotic} to arrive at the final conclusions.

\subsubsection*{Proof of Theorem 2:}

According to the formula \eqref{ITThat_W_adj},
\begin{align*}
	&\widehat{ITT}_W^{adj}-p_{co}^{samp}\\
	=&\widehat{ITT}_W-p_{co}^{samp}-\frac{1}{n_1}\sum_{i:Z_{i}=1}\widehat{\boldsymbol{\beta}}_{W_1}\bm{x}_i^*+\frac{1}{n_0}\sum_{i:Z_{i}=0}\widehat{\boldsymbol{\beta}}_{W_0}\bm{x}_i^*.\\
\end{align*}
Theorem \ref{consistency_CRE} has established $\widehat{ITT}_W-p_{co}^{samp}=o_p(1)$. Denote $S^{-2}_{\bm{x}_z}$ as the inverse of $\boldsymbol{x}_i$'s covariance matrix in treatment arm $z$ and $S^{-2}_{\bm{x}_z}=L^{\top}L$ as its Cholesky decomposition for $z=0,1$. Since $\widehat{\boldsymbol{\beta}}_{W_z}$ is the coefficient vector in linear projection, we have $(\widehat{\boldsymbol{\beta}}_{W_z}L^{-1}(L^{-1})^{\top}\widehat{\boldsymbol{\beta}}_{W_z}^{\top})\leqslant S^2_{W_z}$ for $z=0,1$. This together with the Cauchy-Schwarz inequality imply that for $z=0,1$
\begin{equation}
	\label{bound_on_projection}
	\begin{aligned}
		&\left(\widehat{\boldsymbol{\beta}}_{W_z}(\overline{\boldsymbol{x}}_z-\overline{\boldsymbol{x}})\right)^2\\=&\left(\widehat{\boldsymbol{\beta}}_{W_z}L^{-1}(L^{\top})^{-1}S^{-2}_{\bm{x}_z}(\overline{\boldsymbol{x}}_z-\overline{\boldsymbol{x}})\right)^2\\
		\leqslant&\ S^2_{W_z}(\overline{\boldsymbol{x}}_z-\overline{\boldsymbol{x}})^{\top}S^{-2}_{\bm{x}_z}(\overline{\boldsymbol{x}}_z-\overline{\boldsymbol{x}}).
	\end{aligned}
\end{equation}
Note that $S^2_{W_z}$, $z=0,1$, are bounded. We can apply Lemma A5 in \cite{li2020rerandomization} to the right hand side of \eqref{bound_on_projection} to know it is an $o_p(1)$ term. This implies $\widehat{ITT}_W^{adj}-p_{co}^{samp}=o_p(1)+o_p(1)+o_p(1)=o_p(1)$.

For $z=0,1$, $\frac{n-1}{n_z-1}\mathbb{S}_{Y_z}^2\geqslant S_{Y_z}^2$. By assumption, $\mathbb{S}_{Y_z}^2$ has a finite limit and the limit of $n_z/n$ is in $(0,1)$. It follows that $S_{Y_z}^2$ is also bounded. In a similar way, we can show $\widehat{ITT}_Y^{adj}-ITT_Y^{samp}=o_p(1)$. And the final conclusion can be obtained by repeating the last paragraph of the proof of Theorem \ref{consistency_CRE}.

\subsubsection*{Proof of Proposition 3:}

In the proof of Proposition \ref{conservative_rem}, we show that $S^2_{\widehat{A}_z}-S^2_{A_z}=o_p(1)$ and $S_{\widehat{A}_z,\boldsymbol{x}}-S_{A_z,\boldsymbol{x}}=o_p(1)$ under CRE, hence under ReM by Lemma \ref{dominated_converge_in_p}. Consider
\begin{equation}
	\label{canonic}
	\widetilde{Var}(\widetilde{\tau}_{B}^{\prime})=\frac{S_{B_1^{\prime}}^2}{n_1}+\frac{S_{B_0^{\prime}}^2}{n_0},
\end{equation}
where $S_{B_z^{\prime}}^2$ is the estimated variance of $B_i^{\prime}(z)$ defined as
\begin{equation}
	S_{B_z^{\prime}}^2=\frac{1}{n_z-1}\sum_{i:Z_i=z}\left(B_i^{\prime}(z)-\frac{1}{n_z}\sum_{i:Z_i=z}B_i^{\prime}(z)\right)^2
\end{equation}
We can repeat the establishment of Theorem 8, adding our results to Lemma A5 in \cite{li2020rerandomization}, to conclude that $\widehat{V}_{EHW}-\widetilde{Var}(\widetilde{\tau}_{B}^{\prime})=o_p(n^{-1})$ for $z=0,1$.

Also, Lemma A9 and Theorem 6 in \cite{li2020rerandomization} tell asymptotically equivalence of (\ref{canonic}) to $(1-R^2)Var\left(\widetilde{\tau}_A\right)+(\mathbb{S}^2_{01}-\mathbb{S}^2_{01\mid\boldsymbol{x}})/n=(1-R^{2-})Var\left(\widetilde{\tau}_A\right)^{+}_{\bm{x}}$. Thus, we have $\widehat{V}_{EHW}-(1-R^{2-})Var\left(\widetilde{\tau}_A\right)^{+}_{\bm{x}}=o_p(n^{-1})$.

\subsubsection*{Proof of Proposition 4}

Denote $\Omega$ as the $n\times2(K+1)$ design matrix in which the $i$th row is $(1,Z_i, \bm{x}_i^*, Z_i\bm{x}_i^*)$.
Since $\sum_{i=1}^{n}h_i=\operatorname{trace}(\Omega(\Omega^{\top}\Omega)^{-1}\Omega^{\top})=\operatorname{trace}(\Omega^{\top}\Omega(\Omega^{\top}\Omega)^{-1})=2K+2$ and each $h_i\geqslant0$, we have $\lim_{n\rightarrow\infty}h_i=0$. Thus, the scale multipliers $(1-h_i)^{-1}$ for the HC2 variance estimator and $(1-h_i)^{-2}$ for the HC3 variance estimator both converge to $1$ as $n\rightarrow\infty$. We can repeat the establishment of Theorem 8, adding our results to Lemma A5 and Lemma A12 in \cite{li2020rerandomization} that $\widehat{V}_{HC2}$ and $\widehat{V}_{HC3}$ are asymptotically equivalent to \eqref{canonic}. Thus, we have $\widehat{V}_{HCj}-(1-R^{2-})Var\left(\widetilde{\tau}_A\right)^{+}_{\bm{x}}=o_p(n^{-1})$ for $j=2,3$.

\subsection*{B. Details of the Bayesian Approach}

\subsubsection*{B.1 Prior Distribution}
We use the following flat priors for the parameters. We specify the priors for the coefficients as:
\begin{equation*}
	\begin{aligned}
		(\gamma_{00},\bm{\gamma}_0^{\top})^{\top} & \sim N\left(\bm{0}, 100\mathbb{I}_{K+1}\right),\\
		(\gamma_{10},\bm{\gamma}_1^{\top})^{\top} & \sim N\left(\bm{0}, 100\mathbb{I}_{K+1}\right),\\
		(\beta_{0},\bm{\beta}^{\top})^{\top} & \sim N\left(\bm{0}, 100\mathbb{I}_{K+1}\right),\\
        \alpha & \sim N^+(0,100),
	\end{aligned}
\end{equation*}
where $\mathbb{I}_{K+1}$ is a $(K+1)\times (K+1)$ identity matrix, and $N^+$ denotes a normal distribution truncated over the positive part of real line.

We reparameterize $\sigma_0^2$, $\sigma_1^2$, $\pi_0$ and $\pi_1$ in the covariance matrix for the error terms. Let $\pi_{0 \mid e}$ and $\sigma_{0 \mid e}^2$ denote the population regression coefficient and error variance in a regression of $\varepsilon_{i0}$ on $e_i$, i.e., $\pi_{0\mid e}=\pi_0$ and $\sigma_{0\mid e}^2=\sigma_0^2-\pi_0^2$. Similarly, let $\pi_{1\mid e}=\pi_1$ and $\sigma_{1\mid e}^2=\sigma_1^2-\pi_1^2$.
We specify the following priors:
\begin{equation*}
\begin{aligned}
		\pi_{0 \mid e} & \sim N(0, 100), \\
		\pi_{1 \mid e} & \sim N(0, 100), \\
		\sigma_{0 \mid e}^2 & \sim IG\left(0.01, 0.01\right), \\
		\sigma_{1 \mid e}^2 & \sim IG\left(0.01, 0.01\right).
\end{aligned}
\end{equation*}
Here $IG(0.01,0.01)$ refers to a inverse Gamma distribution with shape 0.01 and scale 0.01.

\subsubsection*{B.2 Gibbs Sampling Algorithm}

We use a Gibbs sampling algorithm with data augmentation to iteratively sample the parameters and the latent variables $(L_i(0),L_i(1))$ from their full conditional distributions.

{\bf Step 1: Sampling the covariance matrix}

For $i=1,\cdots,n$, let $e_i=L_i(0)-\beta_0-\bm{\beta}^{\top}\bm{x}_i$.  For individuals with $W_i=0$, let $\varepsilon_{i0}=Y_i^{obs}-\gamma_{00}-\bm{\gamma}_0^{\top}\bm{x}_i$. For individuals with $W_i=1$, let $\varepsilon_{i1}=Y_i^{obs}-\gamma_{10}-\bm{\gamma}_1^{\top}\bm{x}_i$.

Since the prior distribution of $\sigma_{0\mid e}^2$ is a inverse gamma distribution, and we have $\varepsilon_{i0}\sim N(\pi_{0\mid e}e_i,\sigma_{0\mid e}^2)$ for individuals with $W_i=0$, the conditional distribution of $\sigma_{0\mid e}^2$ is also a inverse gamma distribution:
\begin{equation*}
\sigma_{0\mid e}^2\sim IG(a_0,b_0).
\end{equation*}
Here
\begin{equation*}
\begin{aligned}
a_0&=0.01+\sum_{i=1}^n (1-W_i)/2,\\
b_0&=0.01+\sum_{i:\ W_i=0}(\varepsilon_{i0}-\pi_{0\mid e}e_i)^2/2.
\end{aligned}
\end{equation*}
The conditional distribution of $\pi_{0\mid e}$ is
\begin{equation*}
	\pi_{0\mid e}\sim N(\mu_0, A_0),
\end{equation*}
where
\begin{equation*}
\begin{aligned}
\mu_0&=\left(\sum_{i:\ W_i=0}e_i^2/\sigma_{0\mid e}^2+0.01\right)^{-1}\sum_{i:\ W_i=0}e_i\varepsilon_{i0}/\sigma_{0\mid e}^2,\\
A_0&=\left(\sum_{i:\ W_i=0}e_i^2/\sigma_{0\mid e}^2+0.01\right)^{-1}.
\end{aligned}
\end{equation*}

Similarly, the conditional distribution of $\sigma_{1\mid e}^2$ is
\begin{equation*}
\sigma_{1\mid e}^2\sim IG(a_1, b_1),
\end{equation*}
where
\begin{equation*}
\begin{aligned}
a_1&=0.01+\sum_{i=1}^n W_i/2,\\
b_1&=0.01+\sum_{i:\ W_i=1}(\varepsilon_{i1}-\pi_{1\mid e}e_i)^2/2.
\end{aligned}
\end{equation*}
And the conditional distribution of $\pi_{1\mid e}$ is
\begin{equation*}
	\pi_{1\mid e}\sim N(\mu_1, A_1),
\end{equation*}
where
\begin{equation*}
	\begin{aligned}
		\mu_1&=\left(\sum_{i:\ W_i=1}e_i^2/\sigma_{1\mid e}^2+0.01\right)^{-1}\sum_{i:\ W_i=1}e_i\varepsilon_{i1}/\sigma_{1\mid e}^2,\\
		A_1&=\left(\sum_{i:\ W_i=1}e_i^2/\sigma_{1\mid e}^2+0.01\right)^{-1}.
	\end{aligned}
\end{equation*}

{\bf Step 2: Sampling regression coefficients for $Y_i(0)$ and $Y_i(1)$}

Since the prior distribution of $(\gamma_{00},\bm{\gamma}_0^{\top})^{\top}$ is a normal distribution, and we have $Y_i^{obs}-\gamma_{00}-\bm{\gamma}_0^{\top}\bm{x}_i\sim N(\pi_{0\mid e}e_i,\sigma_{0\mid e}^2)$ for individuals with $W_i=0$, the full conditional distribution of $(\gamma_{00},\bm{\gamma}_0^{\top})^{\top}$ is a normal distribution:
\begin{equation*}
	(\gamma_{00},\bm{\gamma}_0^{\top})^{\top} \sim N\left(\bm{c}_0, \mathbb{C}_{0}^{-1}\right),
\end{equation*}
where
\begin{displaymath}
	\mathbb{C}_{0}=0.01\mathbb{I}_{K+1}+\sum_{i:\ W_i=0} \widetilde{\bm{x}}_{i} \widetilde{\bm{x}}_{i}^{\top} \quad \text { and } \quad \mathbb{C}_{0}\bm{c}_0=\sum_{i:\ W_i=0} \widetilde{\bm{x}}_{i} \widetilde{y}_{i},
\end{displaymath}
with
\begin{equation*}
	\begin{aligned}
		\widetilde{\bm{x}}_{i} &=\left(1, \bm{x}_{i}^{\top}\right)^{\top} / \sigma_{0 \mid e}, \\
		\widetilde{y}_{i} &=\left(Y_i^{obs}-\pi_{0\mid e}e_i\right) / \sigma_{0 \mid e}.
	\end{aligned}
\end{equation*}

The full conditional distribution of $(\gamma_{10},\bm{\gamma}_1^{\top})^{\top}$ can be derived similarly. After sampling regression coefficients, we also update $\varepsilon_{i 0}$ and $\varepsilon_{i 1}$ according to their definitions.

{\bf Step 3: Sampling regression coefficients for $L_i(0)$ and $L_i(1)$}

Let $l_i=L_i(Z_i)$. Let $\pi_{e \mid 0}$ and $\sigma_{e \mid 0}^2$ denote the population regression coefficient and error variance in a regression of $e_i$ on $\varepsilon_{i0}$, i.e., $\pi_{e\mid 0}=\pi_0/\sigma_0^2$ and $\sigma_{e\mid 0}^2=1-\pi_0^2/\sigma_0^2$. Similarly, let $\pi_{e\mid 1}=\pi_1/\sigma_1^2$ and $\sigma_{e\mid 1}^2=1-\pi_1^2/\sigma_1^2$.

Since the prior distribution of $(\alpha,\bm{\beta}^{\top})^{\top}$ is a normal distribution (with the first dimension being truncated), $l_i-\alpha Z_i-\bm{\beta}^{\top}\bm{x}_i\sim N(\pi_{e\mid 0}\varepsilon_{i0},\sigma_{e\mid 0}^2)$ for individuals with $W_i=0$, $l_i-\alpha Z_i-\bm{\beta}^{\top}\bm{x}_i\sim N(\pi_{e\mid 1}\varepsilon_{i1},\sigma_{e\mid 1}^2)$ for individuals with $W_i=1$, the full conditional distribution of $(\alpha,\bm{\beta}^{\top})^{\top}$ is a normal distribution:
\begin{equation*}
	(\alpha,\bm{\beta}^{\top})^{\top} \sim N\left(\bm{d}_0, \mathbb{D}_{0}^{-1}\right)
\end{equation*}
with the first dimension being truncated over the positive part of the real line. Here
\begin{equation*}
	\mathbb{D}_{0}=0.01\mathbb{I}_{K+2}+\sum_{i=1}^n \widetilde{\widetilde{\bm{x}}}_{i} \widetilde{\widetilde{\bm{x}}}_{i}^{\top} \quad \text { and } \quad \mathbb{D}_{0}\bm{d}_0=\sum_{i=1}^n \widetilde{\widetilde{\bm{x}}}_{i} \widetilde{l}_{i},
\end{equation*}
where
\begin{equation*}
	\begin{aligned}
		\widetilde{\widetilde{\bm{x}}}_{i} &=\left(1,Z_{i}, \bm{x}_{i}\right)^{\top} / \sigma_{e \mid 0} \\
		\widetilde{l}_{i} &=\left(l_{i}-\pi_{e\mid 0}\varepsilon_{i0}\right) / \sigma_{e \mid 0}
	\end{aligned}
\end{equation*}
if $W_i = 0$ and
\begin{equation*}
	\begin{aligned}
		\widetilde{\widetilde{\bm{x}}}_{i} &=\left(1, Z_{i}, \bm{x}_{i}\right)^{\top} / \sigma_{e \mid 1} \\
		\widetilde{l}_{i} &=\left(l_{i}-\pi_{e\mid 1}\varepsilon_{i1}\right) / \sigma_{e \mid 1}
	\end{aligned}
\end{equation*}
if $W_i = 1$.

{\bf Step 4: Sampling $l_i=L_i(Z_i)$}

For individuals with $W_i=0$, the latent variable $l_i$ follows a truncated normal distribution
\begin{equation*}
		l_i \sim N^-(\beta_0+\alpha Z_{i} + \bm{\beta}^{\top}\bm{x}_{i}+\pi_{e \mid 0}\epsilon_{i0}, \sigma_{e \mid 0}^2),
\end{equation*}
where $N^-$ denotes a normal distribution truncated over the nonpositive part of real line.

For individuals with $W_i=1$, the latent variable $l_i$ follows a truncated normal distribution
\begin{equation*}
		l_i \sim N^+(\beta_0+\alpha Z_{i} + \bm{\beta}^{\top}\bm{x}_{i}+\pi_{e \mid 1}\epsilon_{i1}, \sigma_{e \mid 1}^2).
\end{equation*}

\subsubsection*{B.3 Imputing the Missing Values}

According to the expression of the $h$th posterior draw of $\tau_{CACE}^{samp}$ in \eqref{tau_postsamp}, we only need to impute the indicator of being a complier $I_{i,co}^{(h)}$ and the missing outcome $Y_i^{mis}$ for each complier.  For ease of presentation, we omit the superscript $(h)$ that indicates the $h$th posterior draw.

\noindent {\bf Step 1: Imputing $I_{i,co}$}

Only units with $W_i^{obs}=Z_i$ can possibly be compliers.  We first consider units with $W^{obs}_i = Z_i = 0$.

Each unit can be a complier (with $W_i(1)=1$ and $W_i(0)=0$) or a never-taker (with $W_i(1)= W_i(0)=0$).  The probability of being a complier given the observed data and the parameters is
	\begin{equation}
		\label{Wmis_Z0}
		\begin{aligned}
			&Pr\left(G_i=co|W_{i}^{obs},Y_{i}^{obs},Z_i,\bm{x}_i,\bm{\Theta}\right)\\
=&Pr\left(W_{i}^{mis}=1-Z_i|W_{i}^{obs},Y_{i}^{obs},Z_i,\bm{x}_i,\bm{\Theta}\right)\\
			=&\frac{Pr\left(W_i(1)=1, W_i(0)=0|Y_{i}^{obs}, \bm{x}_i,\bm{\Theta}\right)}{Pr\left(W_i(1)=1, W_i(0)=0|Y_{i}^{obs}, \bm{x}_i,\bm{\Theta}\right)+Pr\left(W_i(1)= W_i(0)=0|Y_{i}^{obs}, \bm{x}_i,\bm{\Theta}\right)}.
\end{aligned}
\end{equation}

Let $\varepsilon_{i0}=Y_i^{obs}-\gamma_{00}-\bm{\gamma}_0^{\top}\bm{x}_i$.  We have
\begin{equation}
\begin{aligned}
&Pr\left(W_i(1)=1, W_i(0)=0|Y_{i}^{obs}, \bm{x}_i\right)\\
=&Pr\left(L_i(1)>0,L_i(0)\leq 0|Y_{i}^{obs}, \bm{x}_i\right)\\
=&Pr\left(\alpha+\beta_0+\bm{\beta}^{\top}\bm{x}_i+e_i>0,\beta_0+\bm{\beta}^{\top}\bm{x}_i+e_i\leq 0|\varepsilon_{i0}\right)\\
=&Pr\left(-\alpha-\beta_0-\bm{\beta}^{\top}\bm{x}_i<e_i\leq -\beta_0-\bm{\beta}^{\top}\bm{x}_i|\varepsilon_{i0}\right),
\end{aligned}
\end{equation}
and
\begin{equation}
\begin{aligned}
&Pr\left(W_i(1)=0, W_i(0)=0|Y_{i}^{obs}, \bm{x}_i\right)\\
=&Pr\left(L_i(1)\leq 0,L_i(0)\leq 0|Y_{i}^{obs}, \bm{x}_i\right)\\
=&Pr\left(\alpha+\beta_0+\bm{\beta}^{\top}\bm{x}_i+e_i\leq 0,\beta_0+\bm{\beta}^{\top}\bm{x}_i+e_i\leq 0|\varepsilon_{i0}\right)\\
=&Pr\left(e_i\leq -\alpha-\beta_0-\bm{\beta}^{\top}\bm{x}_i|\varepsilon_{i0}\right),
\end{aligned}
\end{equation}
Because
	\begin{equation}
		e_{i}|\varepsilon_{i0} \sim N\left(\pi_0\varepsilon_{i0}/\sigma_{0}^{2}, 1-\pi_0^2/\sigma_0^{2}\right),\label{eq:e_given_eps0}
	\end{equation}
these probabilities can be calculated using normal probabilities.

A unit with $W^{obs}_i = Z_i = 1$ can be a complier (with $W_i(1)=1$ and $W_i(0)=0$) or a always-taker (with $W_i(1)= W_i(0)=1$).  We can similarly sample $W_i^{mis}$.

Finally, we set $I_{i, co}=1$ for units with $W_i^{obs}=Z_i$ and $W_i^{mis}=1-Z_i$, and set $I_{i, co}=0$ for the remaining units.
	
\vspace{0.3cm}
\noindent {\bf Step 2: Imputing $\bm{Y}^{mis}$ for Each Complier}	

We first consider compliers with $W_i^{obs}=Z_i=0$ and $W_i^{mis}=1$.  For each such complier, we have
	\begin{equation}
\begin{aligned}
		Y_i^{obs}=Y_i(0) = \gamma_{00} + \bm{\gamma}_0^{\top}\bm{x}_i + \varepsilon_{i0},\\
		Y_i^{mis}=Y_i(1) = \gamma_{10} + \bm{\gamma}_1^{\top}\bm{x}_i + \varepsilon_{i1}.
\end{aligned}
	\end{equation}
We need to impute $Y_i^{mis}$ conditional on $G_i=co$, $Y_i^{obs}$, $\bm{x}_i$ and the parameters.  We can first impute $\varepsilon_{i1}$ conditional on $G_i=co$, $\varepsilon_{i0}$ and the parameters, and then set
$Y_i^{mis} = \gamma_{10} + \bm{\gamma}_1^{\top}\bm{x}_i + \varepsilon_{i1}$.

Since the covariance structure in \eqref{err} is not fully specified, we need further assumptions to impute the error term $\varepsilon_{i1}$.
The outcome error terms can be written as
	\begin{equation}
		\begin{aligned}
			\varepsilon_{i 0} = \pi_0e_i+\widetilde{\varepsilon}_{i 0},\\
			\varepsilon_{i 1} = \pi_1e_i+\widetilde{\varepsilon}_{i 1},
		\end{aligned}\label{varepsilon}
	\end{equation}
where $\widetilde{\varepsilon}_{i z}\sim{}N\left(0, \sigma_z^{2}-\pi_z^{2}\right)$ for $z=0, 1$.
Let $\rho$ denote the correlation between $\widetilde{\varepsilon}_{i 0}$ and $\widetilde{\varepsilon}_{i 1}$.
Following \cite{Zhang_Johansson_2022}, we take a conservative approach and set $\rho=1$.  With $\rho=1$, we have $\widetilde{\varepsilon}_{i1}=\sqrt{\sigma_1^2-\pi_1^2}/\sqrt{\sigma_0^2-\pi_0^2}\widetilde{\varepsilon}_{i0}$.  Hence
	\begin{equation*}
		\varepsilon_{i 1} = \pi_1e_i+\frac{\sqrt{\sigma_1^2-\pi_1^2}}{\sqrt{\sigma_0^2-\pi_0^2}}(\varepsilon_{i 0}-\pi_0e_i),
	\end{equation*}
We use Rao-Blackwellization to impute $\varepsilon_{i1}$ using its conditional expectation given $G_i=co$, $\varepsilon_{i0}$ and the parameters:
	\begin{equation}
		\varepsilon_{i 1} = \pi_1E\left(e_{i}| G_i=co, \varepsilon_{i 0},\bm{\Theta}\right)+\frac{\sqrt{\sigma_1^2-\pi_1^2}}{\sqrt{\sigma_0^2-\pi_0^2}}\left(\varepsilon_{i 0}-\pi_0E\left(e_{i}| G_i=co, \varepsilon_{i 0},\bm{\Theta}\right)\right).
	\end{equation}
The conditional expectation $E\left(e_{i}| G_i=co, \varepsilon_{i 0},\bm{\Theta}\right)$ can be calculated using the normal distribution in \eqref{eq:e_given_eps0} truncated over the interval from $-\alpha-\beta_0-\bm{\beta}^{\top}\bm{x}_i$ to $-\beta_0-\bm{\beta}^{\top}\bm{x}_i$.

We can similarly impute $Y_i^{mis}$ for each complier with $W_i^{obs}=Z_i=1$ and $W_i^{mis}=0$.

\bibliographystyle{apalike}
\bibliography{Refs}

\begin{thebibliography}{}

\bibitem[Albert and Chib, 1993]{albert1993bayesian}
Albert, J.~H. and Chib, S. (1993).
\newblock Bayesian analysis of binary and polychotomous response data.
\newblock {\em Journal of the American statistical Association},
  88(422):669--679.

\bibitem[Angrist et~al., 1996]{Angrist_etal_1996}
Angrist, J.~D., Imbens, G.~W., and Rubin, D.~B. (1996).
\newblock Identification of causal effects using instrumental variables.
\newblock {\em Journal of the American Statistical Association},
  91(434):444--455.

\bibitem[Bertsimas et~al., 2015]{Bertsimas_etal_2015}
Bertsimas, D., Johnson, M., and Kallus, N. (2015).
\newblock The power of optimization over randomization in designing experiments
  involving small samples.
\newblock {\em Operations Research}, 63(4):868--876.

\bibitem[Ding, 2023]{ding2023first}
Ding, P. (2023).
\newblock A first course in causal inference.
\newblock {\em arXiv preprint arXiv:2305.18793}.

\bibitem[Hirano et~al., 2000]{hirano2000assessing}
Hirano, K., Imbens, G.~W., Rubin, D.~B., and Zhou, X.-H. (2000).
\newblock Assessing the effect of an influenza vaccine in an encouragement
  design.
\newblock {\em Biostatistics}, 1(1):69--88.

\bibitem[Imbens and Rubin, 1997]{imbens1997bayesian}
Imbens, G.~W. and Rubin, D.~B. (1997).
\newblock Bayesian inference for causal effects in randomized experiments with
  noncompliance.
\newblock {\em Annals of Statistics}, 25(1):305--327.

\bibitem[Imbens and Rubin, 2015]{imbens2015causal}
Imbens, G.~W. and Rubin, D.~B. (2015).
\newblock {\em Causal inference in statistics, social, and biomedical
  sciences}.
\newblock Cambridge University Press.

\bibitem[Johansson and Schultzberg, 2020]{Johansson_Schultzberg_2020}
Johansson, P. and Schultzberg, M. (2020).
\newblock Rerandomization strategies for balancing covariates using
  pre-experimental longitudinal data.
\newblock {\em Journal of Computational and Graphical Statistics},
  29(4):798--813.

\bibitem[Johansson and Schultzberg, 2022]{Johansson_Schultzberg_2022}
Johansson, P. and Schultzberg, M. (2022).
\newblock Rerandomization: A complement or substitute for stratification in
  randomized experiments?
\newblock {\em Journal of Statistical Planning and Inference}, 218:43--58.

\bibitem[Kallus, 2018]{Kallus_2018}
Kallus, N. (2018).
\newblock {Optimal a priori balance in the design of controlled experiments}.
\newblock {\em Journal of the Royal Statistical Society. Series B: Statistical
  Methodology}, 80(1):85--112.

\bibitem[Kapelner et~al., 2021]{Kapelneretal_2021}
Kapelner, A., Krieger, A., Sklar, M .and~Shalit, U., and Azriel, D. (2021).
\newblock Harmonizing optimized designs with classic randomization in
  experiments.
\newblock {\em The American Statistican}, 75(2):195--206.

\bibitem[Krieger et~al., 2019]{Kriegeretal_2019}
Krieger, A.~M., Azriel, D., and Kapelner, A. (2019).
\newblock {Nearly random designs with greatly improved balance}.
\newblock {\em Biometrika}, 106(3):695--701.

\bibitem[Lauretto et~al., 2017]{Lauretto_2017}
Lauretto, M.~S., Stern, R.~B., Morgan, K.~L., Clark, M.~H., and Stern, J.~M.
  (2017).
\newblock {Haphazard intentional allocation and rerandomization to improve
  covariate balance in experiments}.
\newblock {\em AIP Conference Proceedings}, 1853(June).

\bibitem[Li and Ding, 2017]{li2017general}
Li, X. and Ding, P. (2017).
\newblock General forms of finite population central limit theorems with
  applications to causal inference.
\newblock {\em Journal of the American Statistical Association},
  112(520):1759--1769.

\bibitem[Li and Ding, 2020]{li2020rerandomization}
Li, X. and Ding, P. (2020).
\newblock Rerandomization and regression adjustment.
\newblock {\em Journal of the Royal Statistical Society Series B: Statistical
  Methodology}, 82(1):241--268.

\bibitem[Li et~al., 2018]{li2018asymptotic}
Li, X., Ding, P., and Rubin, D.~B. (2018).
\newblock Asymptotic theory of rerandomization in treatment--control
  experiments.
\newblock {\em Proceedings of the National Academy of Sciences},
  115(37):9157--9162.

\bibitem[Lin, 2013]{lin2013agnostic}
Lin, W. (2013).
\newblock Agnostic notes on regression adjustments to experimental data:
  Reexamining freedman's critique.
\newblock {\em Annals of Applied Statistics}, 7(1):295--318.

\bibitem[Lopes and Polson, 2014]{lopes2014bayesian}
Lopes, H.~F. and Polson, N.~G. (2014).
\newblock Bayesian instrumental variables: priors and likelihoods.
\newblock {\em Econometric Reviews}, 33(1-4):100--121.

\bibitem[MacKinnon, 2012]{mackinnon2012thirty}
MacKinnon, J.~G. (2012).
\newblock Thirty years of heteroskedasticity-robust inference.
\newblock In {\em Recent advances and future directions in causality,
  prediction, and specification analysis: Essays in honor of Halbert L. White
  Jr}, pages 437--461. Springer.

\bibitem[McNamee, 2009]{McNamee_2009}
McNamee, R. (2009).
\newblock Intention to treat, per protocol, as treated and instrumental
  variable estimators given non-compliance and effect heterogeneity.
\newblock {\em Statistics in Medicine}, 28(21):2639--2652.

\bibitem[Morgan and Rubin, 2012]{morgan2012rerandomization}
Morgan, K.~L. and Rubin, D.~B. (2012).
\newblock Rerandomization to improve covariate balance in experiments.
\newblock {\em The Annals of Statistics}, 40(2):1263--1282.

\bibitem[Rubin, 1978]{Rubin_1978}
Rubin, D.~B. (1978).
\newblock Bayesian inference for causal effects.
\newblock {\em The Annals of Statistics}, 6:34--58.

\bibitem[Shrier et~al., 2014]{Shrier_etal_2014}
Shrier, I., Steele, R.~J., Verhagen, E., Herbert, R., Riddell, C.~A., and
  Kaufman, J.~S. (2014).
\newblock Beyond intention to treat: What is the right question?
\newblock {\em Clinical Trials}, 11(1):28--37.
\newblock PMID: 24096636.

\bibitem[Shrier et~al., 2017]{Shrier_etal_2017}
Shrier, I., Verhagen, E., and Stovitz, S. (2017).
\newblock The intention-to-treat analysis is not always the conservative
  approach.
\newblock {\em The American Journal of Medicine}, 130(7):867--871.

\bibitem[Steele et~al., 2015]{Steele_etal_2015}
Steele, R.~J., Shrier, I., Kaufman, J.~S., and Platt, R.~W. (2015).
\newblock {Simple Estimation of Patient-Oriented Effects From Randomized
  Trials: An Open and Shut CACE}.
\newblock {\em American Journal of Epidemiology}, 182(6):557--566.

\bibitem[Vinokur et~al., 1995]{vinokur1995impact}
Vinokur, A.~D., Price, R.~H., and Schul, Y. (1995).
\newblock Impact of the jobs intervention on unemployed workers varying in risk
  for depression.
\newblock {\em American journal of community psychology}, 23(1):39--74.

\bibitem[Zhang and Johansson, 2022]{Zhang_Johansson_2022}
Zhang, J.~L. and Johansson, P. (2022).
\newblock Model-based bayesian inference under computer assisted
  balance-improving designs.
\newblock {\em Statistics in Medicine}, 41(21):4245--4265.

\end{thebibliography}
\end{document}